**SMART Binary: New Sample Size Planning Resources for SMART Studies**

**with Binary Outcome Measurements**

John J. Dziak[1], Daniel Almirall[2], Walter Dempsey[2], Catherine Stanger[3], Inbal Nahum-Shani[2]

[1] Institute for Health Research and Policy, University of Illinois Chicago

[2] Institute for Social Research, University of Michigan

[3] Center for Technology and Behavioral Health, Geisel School of Medicine, Dartmouth College

**Author Note**

This research was supported by awards R01 DA039901, P50 DA039838, P30 DA029926, and R01 DA015186 from the National Institutes of Health. The content is solely the responsibility of the authors and does not necessarily represent the official views of the funding institutions as mentioned above. The authors thank Nick Seewald and Jamie Yap for their helpful advice and input. We have no known conflict of interest to disclose. Correspondence concerning this paper may be sent to John Dziak, *dziak@uic.edu*, or to Inbal Nahum-Shani, *inbal@umich.edu*. The first and last authors contributed equally. The first author gratefully acknowledges the Methodology Center and Edna Bennett Pierce Prevention Center at The Pennsylvania State University, where he did most of the research and writing for this manuscript, and especially thanks Prevention Research Center director Stephanie Lanza for her support.

The analyses reported here were not preregistered. No empirical dataset was used. Simulation code is available at https://github.com/d3lab-isr/Binary_SMART_Power_Simulations

This article has been accepted for publication in Multivariate Behavioral Research, published by Taylor & Francis.

# Abstract

Sequential Multiple-Assignment Randomized Trials (SMARTs) play an increasingly important role in psychological and behavioral health research. This experimental approach enables researchers to answer scientific questions about how to sequence and match interventions to the unique, changing needs of individuals. A variety of sample size planning resources for SMART studies have been developed in recent years; these enable researchers to plan SMARTs for addressing different types of scientific questions. However, relatively limited attention has been given to planning SMARTs with binary (dichotomous) outcomes, which often require higher sample sizes relative to continuous outcomes. Existing resources for estimating sample size requirements for SMARTs with binary outcomes do not consider the potential to improve power by including a baseline measurement and/or multiple repeated outcome measurements. The current paper addresses this issue by providing sample size simulation code and approximate formulas for two-wave repeated measures binary outcomes (i.e., two measurement times for the outcome variable, before and after receiving the intervention). The simulation results agree well with the formulas. We also discuss how to use simulations to calculate power for studies with more than two outcome measurement occasions. The results show that having at least one repeated measurement of the outcome can substantially improve power under certain conditions.

*Keywords*: Sequential multiple assignment randomized trials (SMARTs), adaptive interventions, binary outcome, power, sample size

## Introduction

Adaptive interventions (also known as dynamic treatment regimens) play an increasingly important role in various domains of psychology, including clinical (Véronneau et al., 2016), organizational (Eden, 2017), educational (Majeika et al., 2020), and health psychology (Nahum-Shani et al., 2015). Designed to address the unique and changing needs of individuals, an adaptive intervention is a protocol that specify how the type, intensity (dose), or delivery modality of an intervention should be modified based on information about the individual's status or progress over time.

As an example, suppose the adaptive intervention in Figure 1 was employed to reduce drug use among youth with cannabis use disorder attending intensive outpatient programs. This example is based on research by Stanger and colleagues (2019) but was modified for illustrative purposes. In this example, youth are initially offered standard contingency management (financial incentives for documented abstinence) with technology-based working memory training (a commercially available digital training program to improve working memory for youth, involving 25 sessions with eight training tasks per session). As part of the intervention, drug use is monitored weekly via urinalysis and alcohol breathalyzer tests over 14 weeks. Second, at week 4, youth who test positive or do not provide drug tests are classified as non-responders and are offered enhanced (i.e., higher magnitude) incentives; otherwise, youth continue with the initial intervention.

This example intervention is "adaptive" because time-varying information about the participant's progress during the intervention (here, response status) is used to make subsequent intervention decisions (here, to decide whether to enhance the intensity of the incentives or continue with the initial intervention). Figure 1 shows how this adaptive intervention can be described with decision rules—a sequence of IF–THEN statements that specify, for each of several decision points (i.e., points in time in which intervention decisions should be made), which intervention to offer under different conditions. Note that this adaptive intervention includes a single tailoring variable: specifically, response status at week 4, measured based on drug tests. Tailoring variables are information used to decide whether and how to intervene (here, whether to offer enhanced incentives or not).

Importantly, an adaptive intervention is not a study design or an experimental design—it is an *intervention design*. Specifically, adaptive intervention is a set of decision rules that can be used to guide practice long after the randomized trial is completed (Collins, 2018; Nahum-Shani & Almirall, 2019). However, in many cases, investigators have scientific questions about how to best construct an adaptive intervention; that is, how to select and adapt intervention options at each decision point to achieve effectiveness and scalability. Sequential multiple assignment randomized trials (SMARTs; Lavori & Dawson, 2000; Murphy, 2005) are increasingly employed in psychological research to empirically inform the development of adaptive interventions (for a review of studies see Ghosh et al., 2020). A SMART is an *experimental design* that includes multiple stages of randomizations and that can be used to provide information for choosing potential adaptive interventions. Each stage is intended to provide data for use in addressing questions about how to intervene and under what conditions at a particular decision point. A SMART is not itself an adaptive intervention; instead, multiple potential adaptive interventions are embedded in the SMART, and the SMART provides randomized evidence about which are likely to be more or less effective.

Consider the working memory training SMART in Figure 2, which was designed to collect data to empirically inform the development of an adaptive intervention for youth with cannabis use disorders (Stanger et al., 2019). This trial was motivated by two questions: in the context of a 14-week contingency management intervention, (a) is it better to initially offer a technology-based intervention that focuses on improving working memory or not? and (b) is it better to enhance the magnitude of incentives or not for youth who do not respond to the initial intervention? These questions concern how to best intervene at two decision points—the first is at program entry and the second is at week four. Hence, the SMART in Figure 2 includes two stages of randomizations, corresponding to these two decision points. Specifically, at program entry youth with cannabis use disorders were provided a standard contingency management intervention and were randomized to either offer working memory training or not. Drug use was monitored weekly via urinalysis and alcohol breath tests over 14 weeks. At week 4, those who were drug positive or did not provide drug tests were classified as early non-responders and were re-randomized to either enhanced incentives or continue with the initial intervention, whereas those who were drug negative were classified as early responders and continued with the initial intervention option (i.e., responders were not re-randomized).

The multiple, sequential randomizations in this example SMART give rise to four "embedded" adaptive interventions (see Table 1). One of these adaptive interventions, labeled "Enhanced working memory training" and represented by cells D+E, was described earlier (Figure 1). Many SMART designs are motivated by scientific questions that concern the comparison between embedded adaptive interventions (Kilbourne et al., 2018; Patrick et al., 2020; Pfammatter et al., 2019). For example, is it better, in terms of abstinence at week 14, to employ the "Enhanced working memory training" adaptive intervention (see Table 1; also represented by cells D, E in Figure 2), or the "Enhanced incentives alone" adaptive intervention (represented by cells A, B in Figure 2)?

Both adaptive interventions offer enhanced incentives to non-responders while continuing the initial intervention for responders, but the former begins with working memory training whereas the latter does not.

The comparison between embedded adaptive interventions is often operationalized using repeated outcome measurements in the course of the trial (Dziak et al., 2019; Nahum-Shani et al., 2020), such as weekly abstinence over 14 weeks measured via weekly drug tests. Repeated outcome measurements in a SMART have both practical and scientific utility (Dziak et al., 2019; Nahum-Shani et al., 2020). They can be leveraged not only to make more precise comparisons of end-of-study outcomes, but also to estimate other quantities, such as area under the curve (AUC; see Almirall et al., 2016), phase-specific slopes, and delayed effects (see Nahum-Shani et al., 2020). Dziak and colleagues (2019) and Nahum-Shani and colleagues (2020) provide guidelines for analyzing data from SMART studies in which the repeated outcome measurements are either continuous or binary. However, although sample size planning resources for SMART studies with numerical repeated outcome measurements have been proposed (e.g., by Seewald et al., 2020), sample size planning resources have yet to be developed for binary repeated outcome measurements. The current paper seeks to close this gap by developing sample size resources for planning SMART studies with binary repeated outcomes measurements.

We begin by reviewing existing sample size planning resources for SMARTs with only an end-of-study binary outcome (i.e., not repeated measurements). We then extend this approach to include a pre-randomization baseline assessment (here called pretest for convenience) and show that this can increase power for comparing adaptive interventions in terms of an end-of-study outcome (i.e., an outcome measured post randomizations which refer to as posttest). In this paper, we provide simulation functions in R to estimate sample size requirements, or power for a given size, in a SMART with binary outcomes and two or more measurement occasions. In the special case of two occasions, we also derive an asymptotic sample size formula which agrees well empirically with the simulation results in the reasonable scenarios considered. We separately consider how to use simulations, constructed appropriately for the SMART context, to calculate power for studies with more than two outcome measurements; an example simulation is given in Appendix 1. It was not practical to derive useful formulas for more than two measurement times. We show by simulations, however, that adding more outcome measurements beyond pretest and posttest may or may not lead to substantial gains in power, depending on the scenario. Nonetheless, these additional measurements may be useful in answering highly novel secondary research questions, such as about delayed effects (see Dziak et al., 2019; Nahum-Shani et al., 2020). It is convenient to start by reviewing the derivation of power and sample size formulas, and then noticing where approximations can reasonably be made and where simulations might be more beneficial.

## Sample Size Planning for Binary SMART

Suppose that in the process of planning the working memory training SMART (Figure 2), investigators would like to calculate the sample size required for comparing the 'enhanced working memory training' and the 'enhanced incentives alone' adaptive interventions (see Table 1). Note that the working memory training SMART is considered a "prototypical" SMART (Ghosh et al., 2020; Nahum-Shani et al., 2022). A prototypical SMART includes two stages of randomization, and the second-stage randomization is restricted to individuals who did not respond to the initial intervention. That is, only non-responders (to both initial options) are re-randomized to second-stage intervention options. More specifically, the first randomization stage involves randomizing all experimental participants to first stage intervention options. Next, response status is assessed. Individuals classified as responders are not re-randomized and typically continue with the initial intervention option. Individuals classified as non-responders are re-randomized to second-stage intervention options. Here, response status is a tailoring variable that is integrated in the SMART by design; that is, this tailoring variable is included in each of the adaptive interventions embedded in this SMART (see Table 1).

### *Notation and Assumptions*

Let $A_1$ denote the indicator for the first-stage intervention options, coded $+1$ for working memory training, or $-1$ for no working memory training; let $R$ denote the response status, coded 1 for responders and 0 for non-responders; and let $A_2$ denote the indicator for the second-stage intervention options among non-responders, coded $+1$ for enhanced incentives and $-1$ for continuing without enhanced incentives. Throughout, we use upper-case letters to represent a random variable, and lower-case letters to represent a particular value of that random variable. Each of the four adaptive interventions embedded in the working memory training SMART (Figure 1) can be characterized by a pair of numbers $(a_1, a_2)$, each $+1$ or $-1$. We write that a participant in a SMART study "follows" or "is compatible with" an adaptive intervention $(a_1, a_2)$ if this participant's first-stage intervention is $a_1$, and if furthermore this participant is either responsive ($R = 1$) to the first-stage intervention, or else is not responsive ($R = 0$) and hence is offered second-stage intervention $a_2$. Notice that this definition includes responders who do not actually receive $a_2$, as long as they did receive $a_1$; the intuition is that they might have received $a_2$ if they had not responded. Thus, unlike in an ordinary randomized trial, the same participant may "follow" more than one of the adaptive interventions being considered;

SMART BINARY

simple statistical approaches to handle this design feature are discussed further by Nahum-Shani and coauthors (2012) and Lu and coauthors (2016).

Let $i=1,…,n$ denote study participants. We assume that, for each $i$, the binary outcomes $Y_{t,i}$ are observed at time points $t = 1, … , T$. Let $R_i(a_1)$ denote the potential outcome of the response status variable (see accessible introduction in Marcus et al., 2012) for person $i$ if that person is offered an adaptive intervention with initial option $a_1$. Let $Y_{t,i}^{(d)}$ or $Y_{t,i}(a_1, a_2)$ denote the potential outcome at time $t$ for person $i$ if offered an adaptive intervention $d$ defined by intervention options $(a_1, a_2)$. It is assumed that if $R_i(a_1) = 1$, then $Y_{t,i}(a_1, -1) = Y_{t,i}(a_1, +1)$, i.e., no one is affected by a part of the adaptive intervention they did not actually receive, although they may still provide information about the effect of an earlier part of the adaptive intervention. Of course, for individuals with $R_i(a_1) = 0$, $Y_{t,i}(a_1, -1)$ need not equal $Y_{t,i}(a_1, +1)$.

For the remainder of the manuscript, we assume that the investigator's goal is to compare a pair of embedded adaptive interventions $d = (a_1, a_2)$ and $d' = (a_1', a_2')$, in terms of outcome probability at end-of-study. We start by reviewing the $T = 1$ case (final, end-of-study outcome only), then extend to $T = 2$ (baseline outcome and final outcome), and then explore $T = 3$ via simulations, using a flexible method that also allows for higher $T$. We assume for most of the paper that the logit link is being used, and that the estimand of interest $\Delta$ is the log odds ratio of the end-of-study outcome between a pair of adaptive interventions. Throughout, we assume that the investigator wishes to choose a sample size $n$ to achieve adequate power to test the null hypothesis $\Delta = 0$. Similar to Kidwell and colleagues (2019) and Seewald and colleagues (2020), we assume that the pair of embedded adaptive interventions being compared differs in at least the first-stage intervention option $A_1$. We also assume that there are no baseline covariates are being adjusted for. In general this is a conservative assumption because adjusting for baseline covariates sometimes improves power and usually does not worsen it (Kidwell et al., 2018).

Recall that the asymptotic sampling variance of a parameter is inversely proportional to the sample size. Across a very wide range of models, the required sample size $n$ to test a null hypothesis $\Delta = 0$ with power $q$ and two-sided level $\alpha$ can be written as

$$n \geq \left(z_q + z_{1-\alpha/2}\right)^2 \frac{\sigma_\Delta^2}{\Delta^2} \qquad (1)$$

where $z_q = \Phi^{-1}(q)$ is the normal quantile corresponding to the desired power, $\Delta$ is the parameter of interest, and $\sigma_\Delta^2$ is a quantity such that for a given sample size $n$, $\mathrm{Var}(\hat{\Delta}) = \sigma_\Delta^2/n$ is its sampling variance; see Derivation 1 in the Appendix 2. The main challenge is to find a formula for $\sigma_\Delta^2$ which fits the model and design of interest, and which can be calculated from intuitively interpretable quantities, for which reasonable guesses could be elicited from a subject matter expert. In this paper we assume that the parameter of interest is the log odds ratio between outcomes for a comparison of two embedded adaptive interventions differing at least in first intervention option. That is, the null hypothesis is $\Delta = 0$ where

$$\Delta = \mathrm{logit}\left(\mu^{(d)}\right) - \mathrm{logit}\left(\mu^{(d')}\right) = \log\left(\frac{\frac{\mu^{(d)}}{1 - \mu^{(d)}}}{\frac{\mu^{(d')}}{1 - \mu^{(d')}}}\right),$$

where $\mu^{(d)} = E\left[Y^{(d)}\right] = P\left[Y^{(d)} = 1\right]$ be the expected value of the binary end-of-study outcome for a participant who follows embedded adaptive intervention $d$. Other quantities of interest, such as the probability ratio, are also possible.

***Parameters Required for Calculating Sample Size***

SMART BINARY

Even after the parameter of interest has been defined and a proposed true value for it has been elicited, more information is still needed to estimate a sample size requirement. These pieces of information could be described as nuisance parameters, although some may be of secondary research interest in their own right. Specifically, let $r_d = E(R^{(d)} = 1)$ be the probability that an individual given adaptive intervention $d$ will be a responder. We assume that $r_d$ depends only on $a_1$ and not on $a_2$, because the second-stage intervention is not assigned until after response status is assessed, but it is still convenient to use the $d$ subscript, with the understanding that $r_d$ and $r_{d'}$ will be the same for adaptive interventions having the same $a_1$. In Appendix 2, we also make consistency assumptions that imply that $\mu^{(d)} = P(Y^{(d)} = 1 | A_1 = a_1, A_2 = a_2)$ and $r_d = P(R = 1 | A_1 = a_1)$. $\mu^{(d)}$ is taken marginally over $R$, representing the overall average success probability for nonresponders who receive the $a_1$ and $a_2$ intervention stages as well as for responders who receive $a_1$ only. Thus, it is not the same as the mean response only of individuals who are observed to receive both $a_1$ and $a_2$.

Let $\psi^{(d0)} = P(Y^{(d)} = 1 | R^{(d)} = 0)$ and $\psi^{(d1)} = P(Y^{(d)} = 1 | R^{(d)} = 1)$ denote the end-of-study outcome probabilities for non-responders and responders, respectively, given intervention and response status. These parameters represent expected values which are conditional on $R$. These parameters can be elicited from investigators by asking them to specify the hypothesized probabilities that $Y$=1 in the six cells A-F in Figure 2. For adaptive intervention $d = (a_1, a_2)$, $\psi^{(d0)}$ corresponds to the probability that $Y$=1 for someone who did not respond to first-stage intervention option $a_1$ and was then offered second-stage intervention option $a_2$. Also, $\psi^{(d1)}$ corresponds to the probability that $Y$=1 for someone who responded to $a_1$. Because responders are not affected by intervention option $a_2$, $\psi^{(d1)}$ is equal for any two adaptive interventions having the same $a_1$, although $\psi^{(d0)}$ is potentially different for each intervention.

SMART BINARY

Although $\psi^{(d1)}$ in particular is dependent on only the first phase (first component) of $d$, it is still convenient to apply the shorthand superscript $d$ here instead of $a_1(d)$, because the adaptive intervention as a whole is assumed to be the target of inference in the analysis.

In the next section, we discuss two options for calculating sample size. The first option requires eliciting hypothetical values of the $\psi^{(d0)}$ and $\psi^{(d1)}$ parameters, which are the end-of-study outcome probabilities *conditional on both the intervention options and response status*. The second option requires eliciting hypothetical values of the $\mu^{(d)}$ parameters, which are the end-of-study outcome probabilities given the embedded adaptive interventions; these probabilities are *conditional only on the intervention options* and are marginal over (i.e., average across levels of) response status.

### *Sample Size Requirements for Posttest Only: A Single Measurement time*

Let $V_d = E\left(\left(Y^{(d)} - \mu^{(d)}\right)^2\right)$ be the variance of $Y^{(d)}$, marginal over $R$. Thus $V_d$ equals $\mu^{(d)}\left(1 - \mu^{(d)}\right)$ because $Y^{(d)}$ is a binary outcome. Also, let $V_{d0} = \mathrm{E}\left(\left(Y^{(d)} - \mu^{(d)}\right)^2 | R = 0\right)$ and $V_{d1} = \mathrm{E}\left(\left(Y^{(d)} - \mu^{(d)}\right)^2 | R = 1\right)$ be the expected squared conditional residuals from the marginal expected outcome for a non-responder or responder, respectively, who follows embedded adaptive intervention $d$. By standard consistency assumptions (see Appendix 2), $V_{d0}$ can also be written as $\mathrm{E}\left((Y - \mu)^2 | A_1 = a_1(d), R = 0, A_2 = a_2(d)\right)$, and $V_{d1}$ can also be written as $\mathrm{E}((Y - \mu)^2 | A_1 = a_1(d), R = 1)$, where $a_1(d)$ and $a_2(d)$ are the intervention options defining adaptive intervention $d$. The quantities $V_{d0}$ and $V_{d1}$ can be calculated indirectly from the elicited probabilities, because

SMART BINARY

$$
\begin{aligned}
V_{d0} = & \ \mathrm{E}\left(\left(Y^{(d)} - \mu^{(d)}\right)^2 | R = 0\right) \\
= & \ \mathrm{E}\left(\left(Y^{(d)} - \psi^{(d0)}\right)^2 | R = 0\right) + \mathrm{E}\left(\left(\psi^{(d0)} - \mu^{(d)}\right)^2 | R = 0\right) \\
& + 2\mathrm{E}\left(\left(Y^{(d)} - \psi^{(d0)}\right)\left(\psi^{(d0)} - \mu^{(d)}\right) | R = 0\right) \\
= & \ \psi^{(d0)}\left(1 - \psi^{(d0)}\right) + \left(\psi^{(d0)} - \mu^{(d)}\right)^2 + 0 \\
= & \ \psi^{(d0)}\left(1 - \psi^{(d0)}\right) + r_d^2\left(\psi^{(d1)} - \psi^{(d0)}\right)^2,
\end{aligned}
$$

and similarly

$$
\begin{aligned}
V_{d1} = & \ \mathrm{E}\left(\left(Y^{(d)} - \mu^{(d)}\right)^2 | R = 1\right) \\
= & \ \psi^{(d1)}\left(1 - \psi^{(d1)}\right) + (1 - r_d)^2\left(\psi^{(d1)} - \psi^{(d0)}\right)^2.
\end{aligned}
$$

Hence, $V_{d0}$ and $V_{d1}$ can be interpreted as the variances of $Y^{(d)}$ conditional on $R$=0 or $R$=1, *plus* an extra quantity that can be interpreted as the effect of response status.

These expressions lead to a sample size recommendation for a pairwise comparison of two adaptive interventions differing at least on stage-1 recommendation. Specifically,

$$
n \geq \frac{\left(z_q + z_{1-\frac{\alpha}{2}}\right)^2 \left(\frac{4(1 - r_d)V_{d0} + 2r_d V_{d1}}{V_d^2} + \frac{4(1 - r_{d'})V_{d'0} + 2r_{d'} V_{d'1}}{V_{d'}^2}\right)}{\Delta^2}, \quad (2)
$$

where $\Delta$ is the true log odds ratio between the adaptive interventions.

Appendix 2 describes how we derived the expression above, using standard causal assumptions, from a sandwich covariance formula

$$
\mathrm{Cov}(\widehat{\boldsymbol{\theta}}) = \frac{1}{n}\boldsymbol{B}^{-1}\boldsymbol{M}\boldsymbol{B}^{-1}.
$$

Here $\boldsymbol{B} = E\left(\sum_d w^{(d)} V_d \boldsymbol{x}_d^T \boldsymbol{x}_d\right) = \sum_d V_d \boldsymbol{x}_d^T \boldsymbol{x}_d$ where $\boldsymbol{x}_d$ is the design matrix expressing adaptive intervention $d$, $w^{(d)}$ is the weight of a given individual under adaptive intervention $d$, and

$$\boldsymbol{M} = E\left(\left(\sum_d w^{(d)} V_d^{-1} \boldsymbol{x}_d^T (Y - \mu^{(d)})\right)^{\otimes 2}\right).$$

Note that weights are employed because non-responders are randomized twice (with probability ½ each time) whereas responders are randomized once (with probability ½), so that the former are under represented in the sample mean under a specific embedded adaptive intervention $d$ (i.e., they have ¼ change of folowing $d$ whereas responders have ½ chance). Thus, inverse probability weights are used (i.e., 4 for non-responders and 2 for responders) to correct for this underrepresentation (see details in Nahum-Shani et al., 2012 and Dziak et al., 2019). Because of the definition of the weights, $\boldsymbol{M}$ simplifies to a diagonal matrix with entries

$$4(1 - r_d)V_{d0} + 2r_d V_{d1}.$$

It is assumed that the target contrast can be written as $\boldsymbol{c}^T \boldsymbol{\theta}$ for some vector $\boldsymbol{c}$, where

$$\sigma_\Delta^2 = \frac{1}{n}\text{Var}(\boldsymbol{c}^T \boldsymbol{\theta}) = \boldsymbol{c}^T \text{Var}\left(\frac{1}{n}\boldsymbol{\theta}\right)\boldsymbol{c} = \boldsymbol{c}^T (\boldsymbol{B}^{-1} \boldsymbol{M} \boldsymbol{B}^{-1})\boldsymbol{c}.$$

In the case of the logistic regression model, this would be true for a pairwise log odds ratio. For a pairwise comparison between adaptive interventions $d$ and $d'$, the researcher would set $c_d = +1$, $c_{d'} = -1$, and other entries of $\boldsymbol{c}$ to zero. After some algebra, the sandwich covariance therefore implies Equation (2). Details are given in Appendix 2.

It appears at first that formula (2) requires specifying hypothetical values for all probabilities, both conditional on $R$ and marginal over $R$, because $V_{d0}$ and $V_{d1}$ depend on both sets of probabilities. However, in practice only the conditional probabilities $\psi^{(d0)}$ and $\psi^{(d1)}$ for each adaptive intervention and the response rate need to be specified, because the marginal probabilities can then be computed by expectations: $\mu^{(d)} = (1 - r_d)\psi^{(d0)} + r_d\psi^{(d1)}$. However, although $\mu^{(d)}$ can be computed from $\psi^{(d0)}$, $\psi^{(d1)}$, and $r_d$, additional assumptions would be needed to compute $\psi^{(d0)}$ and $\psi^{(d1)}$ from $\mu^{(d)}$ and $r_d$.

Kidwell and colleagues (2018) provide an alternative formula, which (in terms of our notation) assumes that $V_{d0} \leq V_d$, $V_{d1} \leq V_d$, $V_{d'0} \leq V_{d'}$, and $V_{d'1} \leq V_{d'}$. Under these variance assumptions, the approximate required sample size is

$$n \geq 2 \frac{\left(z_q + z_{1-\alpha/2}\right)^2}{\Delta^2} \left(\frac{2 - r_d}{V_d} + \frac{2 - r_{d'}}{V_{d'}}\right). \qquad (3)$$

Under the further simplifying assumption that the proportion of responders is equal in the two adaptive interventions being compared ($r_d = r_{d'} = r$), expression (2) simplifies to

$$n \geq 2(2 - r) \frac{\left(z_q + z_{1-\alpha/2}\right)^2}{\Delta^2} \frac{1}{V_d + V_{d'}}.$$

The sample size formula above is equivalent to a sample size formula for a two-arm RCT with binary outcome, multiplied by the quantity $2 - r$, which Kidwell and colleagues (2018) interpreted as a design effect. In practice, this formula requires eliciting hypothetical values for the marginal outcome probabilities $\mu_d$ for each adaptive intervention of interest, and the response rate $r$. Based on these parameters, one can calculate the variance $V_d = \mu_d(1 - \mu_d)$ for each adaptive intervention and calculate the log odds ratio $\Delta = (\mu_d/(1 - \mu_d))/(\mu_{d'}/(1 - \mu_{d'}))$.

Both formula (2) and formula (3) require that the proportion of responders be elicited. Kidwell and colleagues (2019) note that setting $r = 0$ provides a conservative upper bound on required sample size, but the resulting approximation is very pessimistic and may lead to an infeasibly high recommendation.

Both formula (2), which we describe here as a conditional-probabilities-based (CPB) formula, and formula (3) which we describe as a marginal-probabilities-based (MPB) formula, have advantages and disadvantages. The marginal formula requires additional assumptions, but then requires fewer parameters to be elicited. Furthermore, the marginal probabilities are related directly to the marginal log odds ratio of interest for comparing embedded adaptive interventions. In other words, since the hypothesis concerns the comparison of two embedded adaptive interventions, it may be more straightforward for many investigators to specify parameters that describe the characteristics of these adaptive intervention, rather than their corresponding cells. However, other researchers may find the conditional probabilities for each cell comprising the adaptive interventions of interest more intuitive to elicit, as they directly correspond to the randomization structure of the SMART being planned. In the following section, we extend both formulas to settings with a baseline measurement of the outcome.

### Sample Size Requirements for Pretest and Posttest: Two Measurement Times

Power in experimental studies can often be improved by considering a baseline (pre-randomization) assessment as well as the end-of-study outcome (see Benkeser et al., 2021; Vickers & Altman, 2001). These are sometimes described as a pretest and posttest; here, we refer to them as $Y_0$ and $Y_1$. The pretest is assumed to be measured prior to the initial randomization, and therefore causally unrelated to the randomly assigned interventions. The pretest could either be included as a covariate, or else could be modeled as a repeated measure in a multilevel model; we assume the latter approach in the sample size derivations. Below we provide formulas that are similar to (2) and (3), but take advantage of additional information from the baseline measurement.

Let $\mu^{(0)} = E(Y_0)$ be the expected value for the baseline measurement of the outcome at the beginning of the study. Here, neither $Y_0$ nor $\mu^{(0)}$ are indexed by adaptive intervention $d$, because $Y_0$ is measured prior to randomization. Let $\mu^{(d)} = E\left(Y_1^{(d)}\right)$ be the expected value for the end-of-study measurement of the outcome for an individual given adaptive intervention $d$. Then

by Derivation 4 in Appendix 2, the approximate required sample size can be written as

$$n = \frac{\left(z_q + z_{1-\alpha/2}\right)^2}{\Delta^2} \boldsymbol{c}^T \boldsymbol{B}^{-1} \boldsymbol{M} \boldsymbol{B}^{-1} \boldsymbol{c} \qquad (4)$$

where the formulas for $\boldsymbol{c}$, $\boldsymbol{B}$, and $\boldsymbol{M}$ are derived in Appendix 2. The derivation comes from a sandwich covariance formula as in the posttest-only case, and follows the general ideas of Lu and colleagues (2016) and Seewald and colleagues (2020). Specifically $\boldsymbol{B} = \sum_d \boldsymbol{X}_d^T \boldsymbol{S}_d \boldsymbol{X}_d$ where

$\boldsymbol{G}_d$ is a $2 \times 2$ diagonal matrix with entries $\text{Var}(Y_0^{(d)})$ and $\text{Var}(Y_1^{(d)})$, $\boldsymbol{R}_d$ is the $2 \times 2$ within-person correlation matrix between $Y_0^{(d)}$ and $Y_1^{(d)}$, and $\boldsymbol{S}_d = \boldsymbol{G}_d^{\frac{1}{2}} \boldsymbol{R}_d^{-1} \boldsymbol{G}_d^{\frac{1}{2}}$. Under some assumptions

(see Appendix 2), $\boldsymbol{M}$ can be approximated by $\sum_d 4(1 - r_d) \boldsymbol{D}_d^T \boldsymbol{V}_d^{-1} \boldsymbol{V}_{d0} \boldsymbol{V}_d^{-1} \boldsymbol{D}_d +$

$\sum_d 2r_d \boldsymbol{D}_d^T \boldsymbol{V}_d^{-1} \boldsymbol{V}_{d1} \boldsymbol{V}_d^{-1} \boldsymbol{D}_d$.

A formula like (4) can be implemented in code but provides little intuitive understanding.

However, under the further assumption that the variance is independent of response status given

adaptive intervention received, equation (4) simplifies to the following:

$$n = \frac{(2 - r)\left(z_q + z_{1-\alpha/2}\right)^2}{\Delta^2} \left( \frac{4 - 3\rho^2}{2V_d} - \frac{\rho^2}{\sqrt{V_d V_{d'}}} + \frac{4 - 3\rho^2}{2V_{d'}} \right). \qquad (5)$$

The key to the simplifications used in deriving (5) is that $\boldsymbol{B}$ and $\boldsymbol{M}$ can each be expressed as an

"arrowhead" matrix, i.e., a matrix which is all zeroes except for the main diagonal, the first row,

and the first column, and therefore can be inverted by simple algebra, using the formula of

Salkuyeh and Beik (2018). Details are given in Appendix 2.

Although in practice, it is very unlikely that variance will be independent of response status, we use this approximation to generate a formula that is more interpretable and accessible.

The performance of this formula is evaluated later in the simulation studies, where the variance and response status are dependent. Expression (4) is again a CPB formula and Expression (5) is a MPB formula. If the pretest provides no information about the posttest, so that $\rho = 0$, then expression (5) simplifies to expression (3), which was the sample size formula of Kidwell and colleagues (2019). In other words, using an uninformative pretest ($\rho = 0$) is approximately the same as ignoring the pretest.

### Beyond Pretest and Posttest: More than Two Measurement Times

For a SMART with more than two measurement times (i.e., more than pretest and posttest), an easily interpretable formula is not possible without making assumptions that would be unrealistic in the binary case. Seewald and colleagues (2020) provide both a general and a simplified sample size formula for comparing a numerical, end-of-study outcome in longitudinal SMARTs. However, the simplified formula relies on the assumption of homoskedasticity across embedded adaptive interventions and measurement occasions, and exchangeable correlation between measurement occasions. In a binary setting, these simplifying assumptions are less realistic because two binary random variables cannot have equal variance unless they also have either equal (e.g., .20 and .20) or exactly opposite means (e.g., .20 and .80). Determining sample size requirements via simulations would be a feasible alternative in this setting (see Appendix 1).

However, if the investigator prefers not to use simulations, then we propose using the two-measurement-occasion formulas as approximations for planning SMARTs with more than two measurement occasions. Simulations shown in Appendix 1 suggest that the resulting sample size estimates would be reasonable. Although taking more measurement occasions into account might provide somewhat higher predicted power, this would depend on the assumed and true correlation structure and the design assumptions of the SMART. The power could also depend on assumptions concerning the shape of change trajectories within the first- and second-stage of the design (e.g., linear, quadratic, etc.), which might become difficult to elicit. Therefore, although more sophisticated power formulas might be developed, they might offer diminishing returns versus a simpler formula or a simulation. In the next section we discuss the use of simulations to calculate power for settings with more than two measurement times and to investigate the properties of the sample size formulas described earlier.

## Simulation Experiments

In order to test whether the proposed sample size formulas work well, it is necessary to simulate data from SMART studies with repeated binary outcome measurements. Furthermore, simulation code can be relatively easily extended to situations in which the simplifying assumptions of the formulas do not apply. Below we discuss two simulation experiments. The first is designed to assess performance of the power formulas. This is done by comparing, for fixed sample sizes, the power estimated based on the sample size formulas to the power calculated from simulations. The second is designed to assess the performance of the sample size formulas as well as to investigate the extent of reduction in required sample size obtainable by taking pretest into account. This is done by comparing, for a fixed target power, estimates of the required sample sizes given by the various formulas to simulated sample size requirements.

### *Simulation Experiment 1: Performance of Power Formulas*

A factorial simulation experiment was performed based on a SMART design with two measurement times. This experiment investigates the ability of the sample size formulas to choose a sample size which is large enough to achieve 0.80 power under specified assumptions. All simulation code is available online at https://github.com/d3lab-isr/Binary_SMART_Power_Simulations or via https://d3lab.isr.umich.edu/software/ . The experiment is designed to answer the following questions: First, do the proposed sample size formulas accurately predict power compared to the power estimated via simulations? Second, how much does the estimated power change by using the CPB approach in Expression (2), versus the MPB approach in Expression (3)? Third, to what extent does using a pretest result in efficiency gains (i.e., higher power for a given sample size) when comparing adaptive interventions based on repeated binary outcome measurements? Fourth, if the pretest is to be used in the model, is there a relative advantage or disadvantage to including the pretest as a covariate (and only the posttest as an outcome), versus modeling both the pretest and the posttest in a repeated measurement model? We used simulations to answer these questions under a scenario with hypothesized true parameters described below.

*Methods*

Data was simulated to mimic a prototypical SMART study, similar to the working memory training SMART in Figure 1. Randomization probabilities were set to be equal (50% each) for first-stage intervention options for each simulated participant, as well as for second-stage intervention options for each simulated non-responder. We assume there are two outcome measurement occasions: a baseline measurement before randomization (pretest), and an end-of-study outcome measurement (posttest). 10,000 datasets were simulated and analyzed per scenario (combination of effect size and sample size).

We assumed that the contrast of interest is the end-of-study log odds of drug use between the "enhanced working memory" $(+1, +1)$ and the "enhanced incentives alone" $(-1, +1)$ adaptive interventions (Table 1). Also, the data were simulated under the assumption of no attrition (study dropout). In practice a researcher should inflate the final estimate of required sample size to protect against a reasonable estimate of attrition probability.

We compared the power predictions obtained by using the different formulas available for $\sigma_\Delta^2$, with simulated power estimates. Specifically, we considered power calculated from expression (1) using the CPB estimates and MPB estimates for $\sigma_\Delta^2$, which would correspond to the sample size recommendations in expressions (3) and (5), respectively. We generated samples of either $n = 300$ or $n = 500$, in which the true correlation structure was either independent ($\rho = 0$) or correlated with correlation $\rho = .3$ or $\rho = .5$. The datasets were simulated using the approach described below.

*Steps in Simulating Datasets*. We first generated a random dummy variable for baseline abstinence $Y_0$ with probability $E(Y_0) = .40$. Next, $A_1$ was randomly assigned to $+1$ or $-1$ with equal probability. Then, $R$ was generated as a random binary variable (0 or 1) such that the log odds of $R = 1$ was set to $-.62 + Y_0 + .5A_1$. The intercept $-.62$ was chosen to give an overall response rate of about 56% in the $A_1 = +1$ arm and 33% in the $A_1 = -1$ arm, or about 45% overall. Thus, we assume that in general most participants are responders, with an advantage to those receiving working memory training. The correlation between $Y_0$ and $R$ was about .23.

Finally, the end-of-study outcome $Y_1$ was generated. For convenience, $A_2$ and $A_1 \times A_2$ were set to have zero effect, and the effect of $A_1$ was set so that the marginal odds ratio between a pair of adaptive interventions differing on $A_1$ would be approximately 1.5, 2, or 3, depending on the condition. These values are within the ranges which would be considered small, medium and large, respectively, by Olivier, May and Bell (2017). The conditional expected value for the final outcome $Y_1$ is given by the model

$$\text{logit}\big(E(Y_1|Y_0, A_1, R, A_2)\big) = \beta_0 + \beta_{Y_0}Y_0 + \beta_{A_1}A_1 + \beta_R R + \beta_{A_2}A_2 + \beta_{A_1A_2}A_1A_2. \quad (6)$$

The values for $\beta_{A_2}$ and $\beta_{A_1A_2}$ were set to zero for simplicity, and the other values were determined by trial and error to give the desired marginal quantities and are provided in Table 2.

*Analysis of Simulated Datasets.* The model was fit using weighted and replicated estimating equations (see Dziak et al., 2019; Lu et al., 2016; Nahum-Shani et al., 2020) with either working independence or working exchangeable correlation. The latter is equivalent here to working AR-1 because there are only two waves (measurement occasions). Three forms of the twowave model were fit separately: an analysis of the posttest adjusted for pretest as a covariate, a repeated measures analysis with working independence, and a repeated measures analysis with working exchangeable correlation. Tests were done at the standard two-sided Type 1 error rate of .05.

*Computation of Marginal Correlation for Formulas.* Although the two-wave power formulas take the marginal pretest–posttest correlation as an input, this parameter was not directly specified in the simulation code, because a simulation requires fully conditional models to be specified. Therefore, for purposes of calculating power via the formula for a given condition, we used the average marginal correlation estimate obtained from applying the weighted and replicated analysis (marginal over R) to the simulated datasets generated for this specific condition.

*Results*

With respect to the first motivating question (do the proposed sample size formulas accurately predict power compared to the power estimated via simulations), the results of the simulations (Table 3) are very encouraging. First, at least under the conditions simulated, the proposed sample size formulas do predict power accurately compared to the power which is estimated via simulations. As would be expected, power is higher when the effects size is higher and/or the sample size is higher.

The second motivating question concerns the extent that the estimated power will change by using the CPB approach in Expression (2) versus the MPB approach in Expression (3). The results indicate that the MPB and the CPB formulas are equivalent in the one-wave (posttest–only) case. However, these formulas differ slightly from each other in the pretest–posttest scenarios, with the MPB approach being slightly conservative, and the CPB approach being sometimes slightly overly optimistic.

The third question motivating this experiment concerns the extent that using a pretest will result in efficiency gains when comparing adaptive interventions. The results indicate that power is often higher when using a pretest–posttest model than with a posttest-only model, although this depends on within-subject correlation. There is no difference in power between these approaches when the pretest–posttest correlation is negligible (0.06) and only a very small difference when the pretest–posttest correlation is small (0.3), but there is a large difference when the pretest–posttest correlation is sizable (0.6). For example, with an odds ratio of 2 and sample size of 200, the one-wave approach has unacceptably low power of 65%, while the two-wave approach has a much better power of 85%.

Finally, the fourth motivating question concerns the relative advantage or disadvantage to including the pretest as a covariate versus as a measurement occasion in a repeated-measurement model. For purposes of calculating power for comparing adaptive interventions, the working independence analysis was found to be exactly equivalent to a posttest-only analysis, and the covariate-adjusted analysis was essentially equivalent to the exchangeable analysis. Therefore, we focus on comparing results for the non-independent repeated-measures analysis versus the posttest-only analysis. Because we found the simulated power with a pretest covariate to be approximately the same as the simulated power with repeated measures, they are represented by the same column under the Two-Waves heading. This near equivalence may result from the intervention options being randomized in the current settings; had there been confounding, the two models might have dealt with it differently, leading to differences in power and accuracy.

### Simulation Experiment 2: Performance of Sample Size Formulas

This simulation was intended to study the ability of the sample size formulas to choose a sample size which is large enough to achieve a specified power (set here to .80) under specified assumptions, but which is not too large to undermine the feasibility of the study. The questions were analogous to the previous three. First, do the proposed sample size formulas give similar sample size predictions to those obtained from simulations? Second, how much does the estimated sample size change by using the CPB sample size formulas (2) and (4) versus the MPB sample size formulas (3) and (5)? Third, to what extent can the required sample size be reduced, under given assumptions, by taking the pretest–posttest correlation into account?

### Method

Ordinarily, Monte Carlo simulations do not directly provide a needed sample size, but only an estimated power for a given sample size. However, by simulating various points of a power curve and interpolating, it is practical to use simulations to approximate the required sample size. We consider the inverse normal (probit) transform of power, $\Phi^{-1}(q)$, to be approximately linearly associated with $N$, based on the form of Equation (1) and the fact that sampling variance is inversely proportional to $N$. That is, we assume $\Phi^{-1}(q) \approx \hat{a} + \hat{b}N$ for some $\hat{a}$ and $\hat{b}$. Therefore, using the same scenarios as in the previous experiment, we perform simulations for several sample sizes in the range of interest and fit a probit model to relate the predicted power to each sample size. The needed sample size is then roughly estimated as $N =$ $(\Phi^{-1}(.80) - \hat{a})/\hat{b}$. 2,000 datasets were simulated and analyzed per effect size scenario, each on a grid of 10 potential sample sizes.

*Results*

The first question motivating this simulation experiment focused on whether the proposed sample size formulas provide similar sample size predictions to those obtained from simulations. Consistent with the results of the first simulation experiment, the results of the current experiment (Table 4) indicate that the formulas approximately agree with each other, and with the simulations, on the required sample size.

The second motivating question concerns the extent that the estimated sample size changes by using the CPB versus the MPB sample size formulas. As in the first simulation experiment, we found the MPB approach and CPB approach to be practically equivalent in the posttest-only case. In the pretest–posttest case, the MPB approach was found to be slightly conservative and the CPB approach was found to be slightly anticonservative, probably making the MPB approach the safer choice.

Finally, the third question motivating this experiment concerned the extent to which the required sample size can be reduced by taking the pretest–posttest correlation into account. The results indicate that taking pretest–posttest correlation into account reduces the required sample size. As would be expected from the previous simulation experiment, results showed that the required sample size for adequate power can be reduced dramatically (possibly by hundreds of participants) by employing a pretest–posttest approach instead of posttest-only.

**Discussion**

The current manuscript addresses an important gap in planning resources for SMART studies by providing new sample size simulation code, as well as approximate asymptotic sample size formulas, for SMARTs with binary outcomes. These sample size resources enable researchers to consider the inclusion of a pretest when calculating sample size requirements for comparing adaptive interventions. Two simulation experiments show that the new formulas perform well under various realistic scenarios. Given the increased uptake of SMART studies in behavioral science (see Ghosh et al., 2020; Nahum-Shani et al., 2022) and the high prevalence of binary outcome data in many domains of psychological and behavioral health research, the proposed sample size formulas have the potential to contribute to the development of adaptive interventions across multiple fields.

Our simulation results show that taking into account the inclusion of a pretest (i.e., the pretest–posttest correlation) in power calculations leads to smaller sample size requirements than comparing end-of-study outcomes (i.e., posttest) alone. While the sample size savings in some scenarios are relatively small, in other scenarios they are quite substantial, making the SMART design more feasible when resources are limited. Overall, these results suggest that when planning SMART studies with binary outcomes, investigators can potentially improve power by including a baseline measurement. This pretest may be included either as a measurement occasion in a repeated measurement model, or as a covariate, with similar power benefits.

The results also indicate that modeling more outcome measurement occasions beyond pretest and posttest may have diminishing returns in terms of power for comparing end-of-study (posttest) outcomes between adaptive interventions. However, intermediate measurements between pretest and posttest may be vital for secondary research questions about other estimands, such as delayed effects, which are not considered here (see Dziak et al., 2019).

Systematic investigation of the extent of efficiency gained per added measurement occasion is needed to better assess the tradeoff between adding measurement occasions versus adding participants to the study in terms of power for a given hypothesis.

For the pretest–posttest case, we provided both simple asymptotic formulas and simulation code. Simulations have the advantage of being more easily adapted to different designs or situations, and do not require as many simplifying approximations as the asymptotic formulas do, although of course both require assumptions about parameter values.

### *Limitations and Directions for Future Research*

Careful consideration of assumptions, preferably with sensitivity analyses, is still important for sample size planning. It would not be reasonable to argue that planning sample size to achieve exactly .80 power (and no more) is the best approach in general. More conservative sample size approaches may provide more capacity to handle unexpected situations such as higher than anticipated attrition. However, in some cases, an unreasonably high estimated sample size requirement would make it difficult to justify the conduct of a study given realistic funding or participant recruitment constraints. Hence, calculating predicted power with as much precision as possible, for a given set of assumptions, is desirable.

In this paper we have used the ordinary Pearson correlation coefficient, even for describing the relationship between binary variables. This is valid and convenient, and it follows the way correlation is operationalized in, for instance, generalized estimating equations (Liang & Zeger, 1986). However, there are other alternative measures available such as tetrachoric correlation (Bonnett & Price, 2005) which could optionally be explored.  One limitation which might be encountered when choosing parameters for simulations is that very high correlations might lead to complete separation (parameter unidentifiability due to frequentist estimates of certain conditional probabilities being at zero or one). This is a limitation of binary data, but it might be avoided in simulations by not specifying very high correlations, and in practice by either simplifying the analysis model or using priors.

This paper has assumed that sample size calculations would be motivated by a primary question involving a pairwise comparison between two adaptive interventions. However, other estimands could be considered in secondary analyses once the data are gathered. Future studies may extend the sample size planning resources provided in this manuscript to accommodate other planned analyses of binary outcome data from a SMART, such as a multiple comparisons with the best adaptive intervention (Artman et al., 2020; Ertefaie et al., 2016).

**Table 1**

**SMART Design Used by Stanger and colleagues (2019)**

| Adaptive Intervention | $A_1$ | $A_2$ | Stage 1 | Response Status | Stage 2 | Cells (Fig. 1) |
|---|---|---|---|---|---|---|
| Enhanced working memory training | 1 | 1 | WMT+CM | Responder | Continue | D,E |
| | | | | Nonresponder | Add EI | |
| Working memory training alone | 1 | -1 | WMT+CM | Responder | Continue | D,F |
| | | | | Nonresponder | Continue | |
| Enhanced incentives alone | -1 | 1 | CM | Responder | Continue | A,B |
| | | | | Nonresponder | Add EI | |
| Standard contingency management | -1 | -1 | CM | Responder | Continue | A,C |
| | | | | Nonresponder | Continue | |

**Note**: WMT = working memory training, CM = contingency management, EI = enhanced incentives.

**Table 2: Data-Generating Parameters for End-of-Study Binary Outcomes in Simulations**

| Scenario | | Parameters of Data-Generating Model | | | | | | |
|---|---|---|---|---|---|---|---|---|
| Pretest–Posttest Corr. | Effect Size (Odds Ratio) | Conditional Regression Parameters in (6) | | | Marginal Expected Values | | | |
| | | $\beta_0$ | $\beta_{Y_0}$ | $\beta_{A_1}$ | $(-,-)$ | $(-,+)$ | $(+,-)$ | $(+,+)$ |
| 0.06 | 1.5 | -0.44 | 0.00 | 0.100 | 0.45 | 0.45 | 0.55 | 0.55 |
| 0.06 | 2 | -0.44 | 0.00 | 0.250 | 0.42 | 0.42 | 0.59 | 0.59 |
| 0.06 | 3 | -0.44 | 0.00 | 0.460 | 0.37 | 0.37 | 0.63 | 0.64 |
| 0.3 | 1.5 | -0.90 | 1.20 | 0.115 | 0.45 | 0.45 | 0.55 | 0.55 |
| 0.3 | 2 | -0.90 | 1.20 | 0.290 | 0.42 | 0.42 | 0.59 | 0.59 |
| 0.3 | 3 | -0.90 | 1.20 | 0.520 | 0.37 | 0.37 | 0.64 | 0.64 |
| 0.6 | 1.5 | -1.55 | 3.00 | 0.220 | 0.45 | 0.45 | 0.55 | 0.55 |
| 0.6 | 2 | -1.55 | 3.00 | 0.450 | 0.41 | 0.41 | 0.58 | 0.58 |
| 0.6 | 3 | -1.55 | 3.00 | 0.780 | 0.37 | 0.37 | 0.64 | 0.64 |

**Note.** The conditional regression parameters refer to Expression (6). For simplicity, $\beta_R$ is set to 1 and $\beta_{A_2} = \beta_{A_1 A_2} = 0$. This leads to an average percentage of responders across arms of 45%, with responder proportions of 56.5% and 33.5% for the $+1$ and $-1$ levels of $A_1$. Because of a small remaining indirect effect of $Y_0$ and $Y_1$ via $R$ (i.e., correlations between pretest, response variable and posttest), the lowest level of correlation considered here is still not exactly zero (about 0.06), despite specifying a zero parameter for the conditional effect of $Y_0$ and $Y_1$.

**Table 3: Predicted and Simulated Power for Fixed Effect Sizes**

| Scenario | | One Wave | | | Two Wave | | |
|---|---|---|---|---|---|---|---|
| | | Predicted Power | | Simulated Power | Predicted Power | | Simulated Power |
| Odds Ratio | Sample Size | MPB | CPB | | MPB | CPB | |
| Pre-post correlation 0.06 | | | | | | | |
| 1.5 | 300 | 0.298 | 0.298 | 0.296 | 0.299 | 0.313 | 0.298 |
| 1.5 | 500 | 0.455 | 0.454 | 0.457 | 0.456 | 0.476 | 0.458 |
| 2 | 300 | 0.665 | 0.664 | 0.667 | 0.666 | 0.692 | 0.667 |
| 2 | 500 | 0.869 | 0.868 | 0.876 | 0.870 | 0.888 | 0.878 |
| 3 | 300 | 0.956 | 0.955 | 0.961 | 0.956 | 0.966 | 0.962 |
| 3 | 500 | 0.997 | 0.997 | 0.999 | 0.997 | 0.998 | 0.998 |
| Pre-post correlation 0.3 | | | | | | | |
| 1.5 | 300 | 0.286 | 0.286 | 0.279 | 0.312 | 0.324 | 0.305 |
| 1.5 | 500 | 0.437 | 0.437 | 0.449 | 0.475 | 0.493 | 0.484 |
| 2 | 300 | 0.672 | 0.671 | 0.677 | 0.716 | 0.737 | 0.722 |
| 2 | 500 | 0.874 | 0.874 | 0.880 | 0.904 | 0.917 | 0.908 |
| 3 | 300 | 0.957 | 0.957 | 0.963 | 0.971 | 0.977 | 0.976 |
| 3 | 500 | 0.997 | 0.997 | 0.999 | 0.999 | 0.999 | 0.999 |
| Pre-post correlation 0.6 | | | | | | | |
| 1.5 | 300 | 0.290 | 0.290 | 0.291 | 0.423 | 0.431 | 0.427 |
| 1.5 | 500 | 0.442 | 0.442 | 0.450 | 0.627 | 0.637 | 0.641 |
| 2 | 300 | 0.649 | 0.649 | 0.652 | 0.831 | 0.840 | 0.848 |
| 2 | 500 | 0.856 | 0.857 | 0.861 | 0.965 | 0.968 | 0.968 |
| 3 | 300 | 0.955 | 0.956 | 0.962 | 0.994 | 0.995 | 0.996 |
| 3 | 500 | 0.997 | 0.997 | 0.998 | 1.000 | 1.000 | 1.000 |

**Notes**. "MPB" = marginal-probabilities-based (expression 3); "CPB" = conditional-probabilities-based (expression 5), In all conditions, the proportion of responders was set to approximately 0.565 given $A_1 = +1$ and 0.336 given $A_1 = -1$; this difference is the reason why the pre-post correlation $\text{Cor}(Y_0, Y_1)$ could not be set to exactly zero. The odds ratio shown is for pairwise comparison of $(+, -)$ to $(-, -)$ adaptive interventions, which is equivalent here to the effect of $A_1$. For simplicity of interpretation, $A_2$ and the $A_1 \times A_2$ interaction were set to have no effect on $Y_1$. The simulated power shown for the two-wave model uses the covariate adjustment approach (pretest as covariate); the repeated measures approach had approximately the same power, or in some conditions about 0.005% higher.

**Table 4**

**Predicted and Approximate Simulated Sample Size Requirements *N* for Fixed Effect Sizes**

| Scenario | One Wave | | | Two Waves | | |
|---|---|---|---|---|---|---|
| | Predicted *N* Required | | Simulated *N* Required | Predicted *N* Required | | Simulated *N* Required |
| Odds Ratio | MPB | CPB | | MPB | CPB | |
| Pre-post correlation 0.06 | | | | | | |
| 1.5 | 1152 | 1154 | 1157 | 1149 | 1087 | 1153 |
| 2 | 414 | 415 | 413 | 413 | 389 | 412 |
| 3 | 176 | 176 | 167 | 175 | 165 | 165 |
| Pre-post correlation 0.3 | | | | | | |
| 1.5 | 1209 | 1211 | 1210 | 1090 | 1040 | 1091 |
| 2 | 408 | 408 | 411 | 368 | 351 | 371 |
| 3 | 175 | 175 | 169 | 159 | 151 | 152 |
| Pre-post correlation 0.6 | | | | | | |
| 1.5 | 1192 | 1191 | 1182 | 753 | 735 | 745 |
| 2 | 430 | 429 | 431 | 277 | 270 | 269 |
| 3 | 176 | 176 | 172 | 118 | 115 | 109 |

**Notes**. "MPB" = marginal-probabilities-based (expression 3); "CPB" = conditional-probabilities-based (expression 5). The data-generating model settings are the same as those used for Table 3.

**Figure 1**

An example adaptive intervention to reduce drug use among youth with cannabis use disorder attending intensive outpatient programs

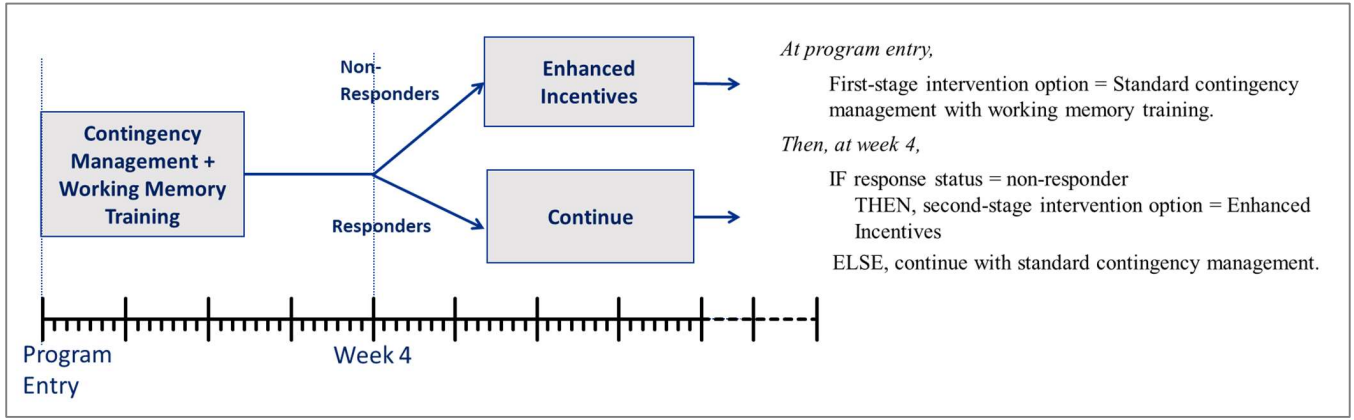

Figure 2

Working Memory Training (WMT) SMART Study

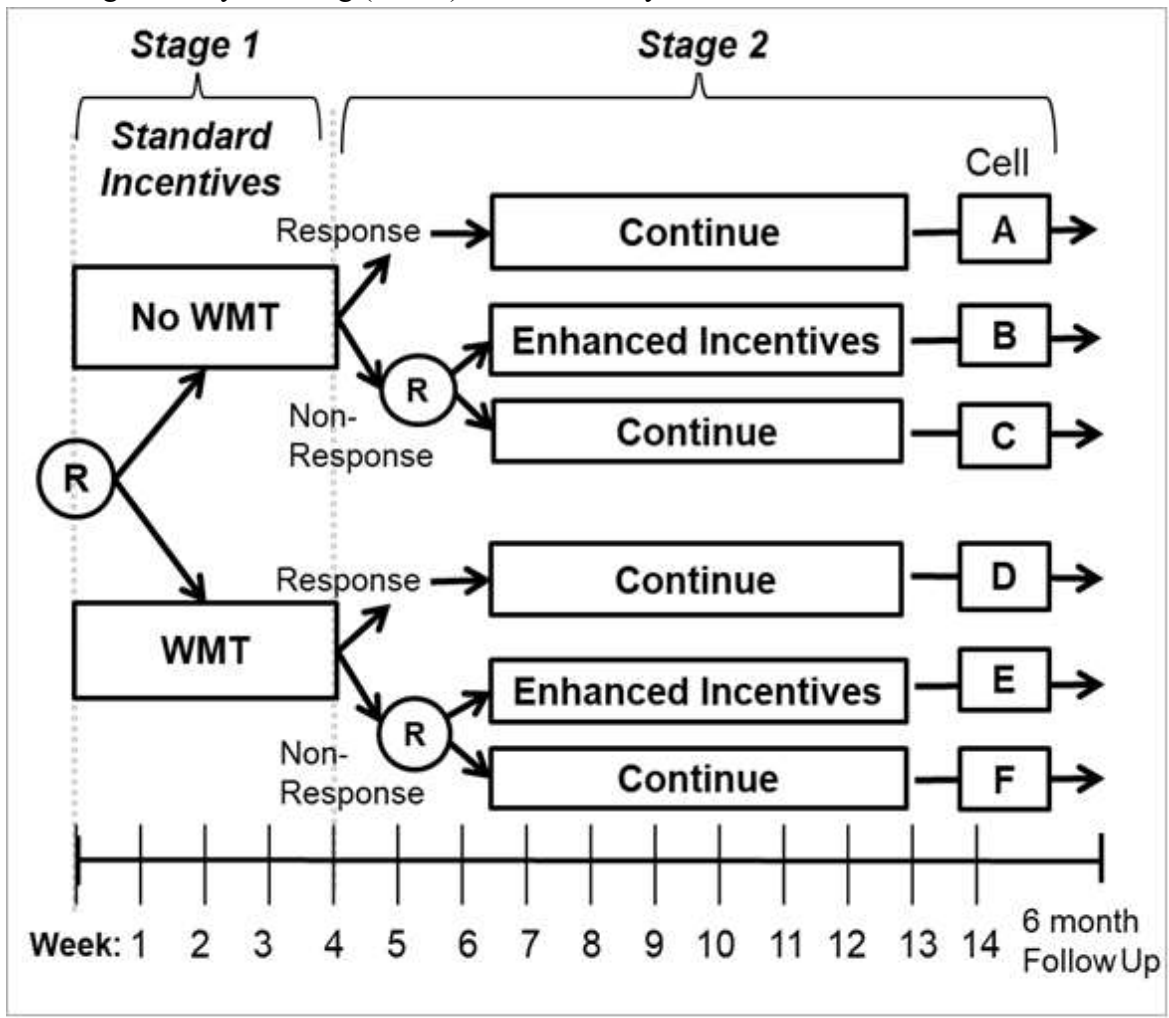

**Notes:** WMT denotes Working Memory Training. Circled R denotes randomization. All participants additionally received contingency management.

# Appendix 1

# Simulation Illustrating Three-Wave Analysis

In this appendix we assume that there are two different follow-up times per participant, $Y_1$ and $Y_2$, instead of a single end-of-study (posttest) outcome $Y_1$, so that there are now three measurement occasions per participant. For simplicity we assume here that both follow-up times occur after the second treatment phase. Therefore, the variables for a given individual in the study would be observed in the following order: pretest $Y_0$, initial randomization $A_1$, tailoring variable $R$, second randomization $A_2$ for nonresponders, first follow-up $Y_1$, second follow-up $Y_2$. This results in a somewhat different and simpler setting compared to that of Seewald (2020), who considered (in the linear modeling case) a mid-study outcome taken about the same time as $R$, and preceding the second randomization.

In the current setting, the second outcome $Y_2$ can potentially depend on any of $Y_0$, $A_1$, $R$, $A_2$, and $Y_1$, making a very wide range of different DAGs and simulation scenarios possible. For simplicity, we chose scenarios in which $Y_2$ depends on $Y_1$ but is conditionally independent of most or all the other preceding variables given $Y_1$.

Specifically, we continue to assume the same parameters used in the "high" correlation setting from Simulations 1 and 2 when simulating $Y_0$ and $Y_1$. We then further assume one of two scenarios for the relationship of $Y_2$ to the preceding variables. In the "no delayed effect" scenario, $Y_2$ depends on the preceding variables only through $Y_1$, and is conditionally independent of all other variables given $Y_1$, as if in a Markov chain. Thus, the effect of $A_1$ on $Y_2$ is mediated entirely by $Y_1$. In the "delayed effect" scenario, $A_1$ has an effect on $Y_2$ which is only partly mediated by $Y_1$.

SMART BINARY

The conditional models used for these two scenarios are as follows. For the "no delayed effect" condition,

$$\text{logit}\big(E(Y_2|Y_0, A_1, R, A_2, Y_1)\big) = \text{logit}\big(E(Y_2|Y_1)\big) = -1.4 + 3Y_1.$$

For the "delayed effect" condition,

$$\text{logit}\big(E(Y_2|Y_0, A_1, R, A_2, Y_1)\big) = \text{logit}\big(E(Y_2|A_1, Y_1)\big) = -1.4 + .275A_1 + .5Y_1.$$

The conditional effect of $Y_1$ on $Y_2$ was set to be weaker in the delayed effect scenario, so that the total effect of $A_1$ on $Y_2$ (i.e., the direct effect conditional on $Y_1$ plus the indirect effect mediated through $Y_1$) would be comparable between scenarios. In particular, the resulting odds ratio for the contrast of interest, still assumed to be $(+1, -1)$ versus $(-1, -1)$, was 3.0 for $Y_1$ and 2.0 for $Y_2$.

We assume that the estimand of interest is comparison of embedded adaptive interventions on the final outcome, where final outcome is interpreted as either the early follow-up $Y_1$ or the later follow-up $Y_2$, in order to compare the simulated power for each. We fit one-wave models to predict $Y_1$ alone or $Y_2$ alone. We also fit two-wave models to predict $Y_1$ or $Y_2$ separately adjusting for $Y_0$, and assuming exchangeable correlation structure (equivalent to AR-1 for the two-wave model). These models only consider two of the measurement occasions available. Finally, we fit three-wave models to predict $Y_2$ adjusting for $Y_0$ and $Y_1$, by applying methods similar to Lu and colleagues (2016) and Dziak and colleagues (2020) and using working assumptions of either independence, AR-1 or exchangeable correlation structure. In the three-wave models, we assumed a separate piecewise linear trajectory from $Y_0$ to $Y_1$ and from $Y_1$ to $Y_2$ for each embedded adaptive intervention.

Each scenario was replicated in 10,000 datasets each for sample sizes $n = 300$ and $n = 500$. Simulated power for each model in each scenario is shown in Table 5. Power for models using $Y_1$ as the final outcome was very high, and much higher than those using $Y_2$ as the final outcome. However, this is not surprising because the effect size for $Y_1$ was also higher. More interesting is the power comparison among the five models for $Y_2$ (the rightmost five columns).

In the no-delayed-effect scenario, power was clearly higher for methods which used information from $Y_0$ to predict $Y_2$ (i.e., "$Y_2$ Adjusted for $Y_0$," working AR-1, and working exchangeable) versus those which ignored $Y_0$ ("$Y_2$ Only" and working independence). However, there was very little additional benefit in using $Y_1$, possibly because $Y_1$ is on the causal chain between $Y_0$ and $A_1$ on the left, and $Y_2$ on the right. Also, as expected, power was higher for a working correlation that approximately fit the data-generating model (AR-1) than one which did not (exchangeable). Although neither structure corresponded exactly to the data-generating model, the exchangeable working structure made the unhelpful assumption that $\mathrm{Corr}(Y_0, Y_1) = \mathrm{Corr}(Y_0, Y_2)$. In contrast, in the delayed effect scenario, it made little difference which model was used. This was presumably because in this scenario $Y_0$ and $Y_1$ had relatively little value for predicting $Y_2$ once $A_1$ was accounted for.

There are many other possible data-generating models that could be explored in a three-wave simulation. For instance, we did not explore power for detecting an effect of $A_2$, or whether power might be different depending on the order and timing of the measurements. However, it appears that at least in some circumstances, a two-wave ($Y_0 \rightarrow Y_2$) model provides about as much benefit as a three-wave model ($Y_0 \rightarrow Y_1 \rightarrow Y_2$) with less complexity, assuming that contrasts in expected values for $Y_2$ are of primary interest. Of course, for more complicated estimands (e.g., for studying whether the effect is delayed), more than two waves would be needed.

**Table 5 (for Appendix 1)**

**Simulated power for different models in three-wave simulation**

| Scenario | | Simulated power for first follow-up $Y_1$, by model | | Simulated power for second follow-up $Y_2$, by model | | | | |
|---|---|---|---|---|---|---|---|---|
| Delayed Effect | Sample Size | $Y_1$ Only | $Y_1$ Adjusted for $Y_0$ | $Y_2$ Only | $Y_2$ Adjusted for $Y_0$ | $Y_2$ Adjusted for $Y_0$ and $Y_1$ | | |
| | | | | | | (Indep.) | (AR-1) | (Exch.) |
| No | 300 | 0.962 | 0.997 | 0.651 | 0.715 | 0.651 | 0.726 | 0.699 |
| No | 500 | 0.998 | 1.000 | 0.859 | 0.907 | 0.859 | 0.909 | 0.893 |
| Yes | 300 | 0.961 | 0.997 | 0.515 | 0.517 | 0.515 | 0.540 | 0.521 |
| Yes | 500 | 0.998 | 1.000 | 0.744 | 0.746 | 0.744 | 0.757 | 0.737 |

**Notes**. In all of these conditions, the average estimated odds ratio for the effect of $A_1$ was set to 3.0 for $Y_1$ and 2.0 for $Y_2$, in terms of the pairwise comparison of $(+, -)$ to $(-, -)$ adaptive interventions, which is equivalent here to the effect of $A_1$. For simplicity of interpretation, $A_2$ and the $A_1 \times A_2$ interaction were set to have no effect. The conditions differ in the relationship of the simulated late follow-up $Y_2$ to the baseline assessment $Y_0$ and initial treatment $A_1$. The simulated decay in effect size over time between $Y_1$ and $Y_2$ is intended to be analogous to that found in many real-world clinical trials.

SMART BINARY

## Appendix 2: Derivations of Formulas

**Derivation of Expression (1)**

The approximate power $q$ to reject the null hypothesis is typically found by setting:

$$q = \Phi\left(\sqrt{\frac{\hat{\Delta}^2}{\mathrm{Var}(\hat{\Delta})}} - z_{1-\frac{\alpha}{2}}\right) \tag{7}$$

This formula can be derived for a generic Wald $Z$-test as follows. By the central limit theorem, for sufficiently large sample size $n$,

$$Z = \frac{\hat{\Delta} - \Delta}{\sqrt{\mathrm{Var}(\hat{\Delta})}}$$

has approximately a standard normal distribution, where $\mathrm{Var}_n(\hat{\Delta})$ is the sampling variance of the estimate $\hat{\Delta}$ given sample size $n$. Henceforth we hide the subscript $n$ and simply write $\mathrm{Var}(\hat{\Delta})$, but this is still implicitly understood to depend on sample size. Consider a two-sided test of whether $\Delta = 0$, and assume without loss of generality that $\Delta > 0$. The null hypothesis is rejected in the correct direction if the test statistic $\hat{\Delta}/\sqrt{\mathrm{Var}(\hat{\Delta})}$ is greater than the critical value $z_{1-\alpha/2} = \Phi^{-1}(1 - \alpha/2)$, that is, the $1 - \alpha/2$ quantile of the standard normal distribution. The probability of rejecting the null hypothesis is then

$$P(Z > z_{1-\alpha/2}) = P\left(\frac{\Delta}{\sqrt{\mathrm{Var}(\hat{\Delta})}} + \frac{\hat{\Delta} - \Delta}{\sqrt{\mathrm{Var}(\hat{\Delta})}} - z_{1-\alpha/2} > 0\right)$$

$$= P\left(\frac{\hat{\Delta} - \Delta}{\sqrt{\mathrm{Var}(\hat{\Delta})}} < \frac{\Delta}{\sqrt{\mathrm{Var}(\hat{\Delta})}} - z_{1-\alpha/2}\right).$$

SMART BINARY

Expression (7) can be rearranged to solve for sample size as follows. Asymptotically, $\text{Var}(\hat{\Delta})$ is inversely proportional to the sample size $n$, so that there is some quantity $\sigma_\Delta^2$ for which we could write $\text{Var}(\hat{\theta}) \approx \frac{1}{n}\sigma_\Delta^2$ for large $n$. Therefore, the needed sample size for a power $q$ can then be expressed as in Expression (1) (as in, e.g., Kidwell et al., 2019; Seewald et al., 2020). Thus, Expression (7) can be rewritten as

$$q = \Phi\left(\sqrt{\frac{\hat{\Delta}^2}{n\sigma_\Delta^2}} - z_{1-\alpha/2}\right).$$

where $\text{Var}(\hat{\Delta}) = \sigma_\Delta^2/n$. Solving this for $n$ leads to Expression (1).

**Derivation of Expression (2)**

Suppose a researcher is considering sample size required for a two-stage restricted (Design II) SMART with weighted and replicated estimating equations (see, e.g., Lu et al., 2016; Kidwell et al., 2019; Seewald et al., 2020), either with a single endpoint or with repeated measures. Suppose further that the contrast for which power is being planned is the comparison of two embedded adaptive interventions (i.e., dynamic treatment regimens), $d$ and $d'$, in terms of end-of-study outcomes. We assume that the difference in effectiveness will be measured by a log odds ratio after fitting a logistic model.

Following the potential outcomes framework, we consider each individual participant to have had, in advance, potential response status $R_i^{(d)}$ and $Y_i^{(d)}$ for each possible adaptive intervention $d = \left((a_1)_d, ((a_2)_d)\right)$ which that participant could receive. As usual, some of these will be unobserved counterfactual outcomes. As a special consequence of the design of the study, $R_i^{(d)}$ actually depends only on $(a_1)_d$ and not $(a_2)_d$. Furthermore, if $R_i^{(d)} = 1$, then $Y_i^{(d)}$ also depends only on $(a_1)_d$ and not $(a_2)_d$, because the second-phase intervention option which the participant would otherwise have received, is not assigned, or at least not revealed or used. It may seem counterintuitive to be making inferences for the effect of $((a_1)_d, ((a_2)_d))$ as a whole, when it is known that the second component is irrelevant to some study participants. However, the analysis is intended to provide generalizable information on which adaptive intervention would be best on average for future patients, considering both responders and nonresponders (Lu et al, 2016; Dziak et al., 2019; Nahum-Shani et al., 2020). When writing expectations, we suppress the $i$ subscript in $R_i^{(d)}$ and $Y_i^{(d)}$ and simply write $E(R^{(d)})$ and $E(Y^{(d)})$.

In the case of a significance test of the log odds ratio, it is assumed that the log odds ratio will be expressed as a linear combination of the coefficients $\boldsymbol{\theta}$ of a longitudinal marginal model with binary outcomes and a logit link, where $\boldsymbol{\mu}^{(d)} = \text{logit}^{-1}(\boldsymbol{X\theta})$. Following Lu and colleagues (2016), the investigator would estimate the coefficients by solving

$$\sum_{i=1}^n \sum_d w_i^{(d)} \boldsymbol{D}_d^T \boldsymbol{V}_d^{-1}(\boldsymbol{Y}_i - \boldsymbol{\mu}^{(d)}(\boldsymbol{\theta})) = \boldsymbol{0} \qquad (8)$$

where $\boldsymbol{V}_d = \text{Cov}(\boldsymbol{Y}^{(d)})$ and $\boldsymbol{D}_d = \partial \boldsymbol{\mu}^{(d)}/\partial \boldsymbol{\theta}$.

As in Kidwell and colleagues (2019) and Seewald and colleagues (2020), we use the weights

$$w_i^{(d)} = 2\mathbb{I}\{A_1 = a_{1d}\}(R + 2(1-R)\mathbb{I}\{A_2 = a_{2NRd}\})$$

where $A_1$ and $A_2$ are effect-coded except that $A_2 = 0$ for responders. We also assume that randomizations will be done with equal probability. As a result, responders are compatible with two of the four embedded adaptive interventions, and have a weight of $1/(1/2) = 2$, while responders are compatible with only one of the four adaptive interventions, and have a weight of $1/(1/4) = 4$. These weights can be thought of as inverse propensity weights, or equivalently as adjustments in a weighting and replication approach (see Lu and colleagues, 2016).

We are assuming the estimand $\Delta$ of interest is a linear combination $\Delta = \boldsymbol{c}^T\boldsymbol{\theta}$ of the regression coefficients. To derive expression (2) from expression (1), it is necessary to have a value for $\sigma_\Delta^2 = n\text{Var}(\boldsymbol{c}^T\widehat{\boldsymbol{\theta}})$. However, $\sigma_\Delta^2$ itself is an abstract quantity which would be impractical to elicit from a clinician or substantive researcher, and would not be reported in literature. Thus, it must be reexpressed in terms of other quantities. Under our assumption that $\Delta = \boldsymbol{c}^T\boldsymbol{\theta}$, we have $\text{Var}(\boldsymbol{c}^T\widehat{\boldsymbol{\theta}}) = \boldsymbol{c}^T\text{Cov}(\widehat{\boldsymbol{\theta}})\boldsymbol{c}$. This means the remaining problem is to re-express $\text{Cov}(\widehat{\boldsymbol{\theta}})$ in terms of quantities that can be easily interpreted and elicited.

Suppose that, as in Kidwell and colleagues (2019), we only consider data from the final measurement time (i.e., no baseline observation). Then Equation (8) becomes

$$\sum_{i=1}^n \sum_d w_i^{(d)} \boldsymbol{D}_d^T V_d^{-1}(Y_i - \mu^{(d)}(\boldsymbol{\theta})) = \boldsymbol{0}.$$

For adaptive intervention $d = (a_1, a_{2NR})$, the mean function is $\mu^{(d)}(\boldsymbol{\theta}) = \text{logit}^{-1}(\boldsymbol{x}_d\boldsymbol{\theta})$ for adaptive-intervention-specific design matrices $\boldsymbol{x}_d$ and adaptive-intervention-specific Jacobian matrices $\boldsymbol{D}_d = \mu^{(d)}(1 - \mu^{(d)})\boldsymbol{x}_d$. Because of the binary outcome, $V_d = \text{E}\left(\left(Y^{(d)}\right)^2\right) - \left(\text{E}\left(Y^{(d)}\right)\right)^2 = \text{E}\left(Y^{(d)}\right) - \left(\text{E}\left(Y^{(d)}\right)\right)^2 = \mu^{(d)} - \left(\mu^{(d)}\right)^2 = \mu^{(d)}(1 - \mu^{(d)})$. This is just the expected variance formula for a Bernoulli (binary) random variate; that is, the dependence of $Y$ on $R$ does not create overdispersion here because 0 and 1 are their own squares, although overdispersion might occur for a more general binomial (event count) $Y$. Thus, assuming $0 < \mu^{(d)} < 1$ for each $d$, Equation (1) simplifies further to

$$\sum_{i=1}^n \sum_d w_i^{(d)} \boldsymbol{x}_d^T(Y_i - \mu^{(d)}(\boldsymbol{\theta})) = \boldsymbol{0}.$$

which resembles an ordinary (independence) weighted logistic regression.

Following Lu and colleagues (2016) and Kidwell and colleagues (2019), we use a sandwich formula $\text{Cov}(\widehat{\boldsymbol{\theta}}) = \frac{1}{n}\boldsymbol{B}^{-1}\boldsymbol{M}\boldsymbol{B}^{-1}$. In order to proceed in computing this sandwich formula, we make three further assumptions. The first two are consistency assumptions (see Seewald et al., 2019; see, e.g., Hernán & Robins, 2020 for a more general discussion of consistency) needed in order for the results of the proposed study to be identified and interpretable. The third is a working assumption (similar to those by Seewald and colleagues, 2020) made in order to allow for a tractable expression for $\sigma_\Delta^2$ in terms of elicitable quantities. Let $d$ be an embedded adaptive intervention defined by intervention options $(a_1, a_2)$.

1. **Consistency assumption** for $R$: The observed response $R$ for a given individual consistent with adaptive intervention $d$ equals that individual's potential outcome $R^{(d)}$. This assumption is required in order for the model quantities to be identifiable. We use the notation $R^{(d)}$ for convenience, but it might be more precise to write $R^{((a_1)_d)}$, because $R^{(d)}$ depends only on the $a_1$ component of adaptive intervention $d$, not the $a_2$ component, which would not even be used except after observing nonresponse. That is, if the embedded adaptive interventions are listed as (+1,+1), (+1,-1), (-1,+1), and (-1,-1), then $R^{(1)} = R^{(2)}$ by construction and $R^{(3)} = R^{(4)}$. Nonetheless, the notation $R^{(d)}$ is less cumbersome than a more precise alternative would be. We write the expectation of $R^{(d)}$ as $r_d = E(R^{(d)})$.

2. **Consistency assumption** for $Y$: The observed outcome $Y$ for a given individual who is consistent with adaptive intervention $d$, and who has response status $R^{(d)}$, equals that individual's potential outcome $Y^{(d)}$. This is required in order for the model quantities to be identifiable. In the current context $Y$ is a binary variable with expected value $\mu^{(d)}$, and can therefore be shown to have variance $\mu^{(d)}(1 - \mu^{(d)})$.

    3. **Cross-world independence assumption** for responders: For two adaptive interventions $d$ and $d'$ with the same initial intervention option $a_1$ but different second intervention options,

$$E\big((Y^{(d)} - \mu^{(d)})(Y^{(d')} - \mu^{(d')})|R^{(d)} = 1\big) = 0.$$

This assumption cannot be checked because responders cannot be consistent with both adaptive interventions $d$ and $d'$ and therefore, for responders, only one of $Y^{(d)}$ or $Y^{(d')}$ can ever be observed. However, it seems reasonable under the assumption that intervention option is randomized. It does not matter whether the variable being conditioned on is written as $R^{(d)}$ or $R^{(d')}$, because the adaptive interventions are assumed to entail the same initial intervention option.

    As in Seewald and colleagues (2020), the bread of the sandwich, i.e., the naïve model-based covariance matrix for the longitudinal regression parameters before adjusting for the special features of the SMART design, is $\boldsymbol{B}^{-1}$, where $\boldsymbol{B}$ is the weighted sum of squares

$$\boldsymbol{B} = E\big(\textstyle\sum_d w^{(d)} V_d^{-1} \boldsymbol{D}_d^T \boldsymbol{D}_d\big).$$

    Recall that responders have a $1/2$ chance of being compatible with a given adaptive intervention (depending only on the randomization of $a_1$) and have a weight of 2, while nonresponders have a $1/4$ chance (depending on the randomization of $a_1$ and $a_2$) and have a weight of 4. Because of this, the expected values of the inverse probability weights $w^{(d)}$ are 1 by the law of iterated expectations. The $V_d$ and $\boldsymbol{D}_d$ are not dependent on observed data but only on the true parameters (recall that we are assuming no baseline covariates) so they can be treated as

SMART BINARY

constants. That is, assuming consistency assumption 1 in section 4.1, and using the weight $w_i^{(d)}$ given in (2), a given cross-product term in (6) simplifies to

$$E(w^{(d)}V_d^{-1}\boldsymbol{D}_d^T\boldsymbol{D}_d) = 2E(I\{A_{1i} = a_1\}RV_d^{-1}\boldsymbol{D}_d^T\boldsymbol{D}_d) + 4E(\{A_{2i} = a_2\}(1 - R)V_d^{-1}\boldsymbol{D}_d^T\boldsymbol{D}_d)$$

$$= 2(\tfrac{1}{2})r_dV_d^{-1}\boldsymbol{D}_d^T\boldsymbol{D}_d + 4(\tfrac{1}{2})(1 - r_d)V_d^{-1}\boldsymbol{D}_d^T\boldsymbol{D}_d = V_d^{-1}\boldsymbol{D}_d^T\boldsymbol{D}_d.$$

Therefore,

$$\boldsymbol{B} = \sum_d V_d^{-1}\boldsymbol{D}_d^T\boldsymbol{D}_d = \sum_d V_d\boldsymbol{x}_d^T\boldsymbol{x}_d.$$

As an aside, it may seem strange that the exponent of $V_d$ in the expression above is positive, not negative, since $\boldsymbol{B}^{-1}$ is the model-based covariance estimate. All else being equal, the higher a $V_d$ is, the smaller the error variance of the effect of interest will be. In that sense, $V_d$ is behaving opposite to how a variance term $\sigma^2$ would behave in a linear model. However, this does not constitute a problem with the model or the estimator; it is instead a result of the basic properties of the logistic link function. Roughly speaking, when the mean (i.e., probability parameter) of a Bernoulli variable is near 0 or near 1, the variance of the observed values will be very small, because most observed values of the variable will be identical (0 or 1, respectively). However, under those circumstances the error variance of the *odds ratio* will be large, because either the odds ratio or its inverse is tending towards infinity. When the probability is instead near 1/2, the variance of the probability estimate is large (because the observed values are roughly evenly distributed between 0's and 1's, none of them near the mean). However, this is the circumstance in which the odds ratio has the lowest variance. In general, for a Bernoulli random variable $x$ with probability $p$, we have $\mathrm{Var}(\bar{x}) = n^{-1}p(1 - p)$, but by the delta method $\mathrm{Var}(\log(\bar{x}/(1 - \bar{x})) = n^{-1}p^{-1}(1 - p)^{-1}$. Intuitively, the equivalent of $\sigma^2$ in logistic regression models is $V^{-1}$, which determines the variance of the estimand of interest (log odds ratio), rather than $V$, which determines the variance of $Y$ itself.

Returning to the expression for $\boldsymbol{B}$, it is still necessary to handle the sum over adaptive interventions $d$. It is most convenient for purposes of derivation to have a specific dummy code (indicator) for each adaptive intervention instead of having dummy or effect codes with a general intercept term. Therefore, with the four embedded adaptive interventions, we can write

$$\boldsymbol{x}_1 = \begin{bmatrix} 1 \\ 0 \\ 0 \\ 0 \end{bmatrix}, \boldsymbol{x}_2 = \begin{bmatrix} 0 \\ 1 \\ 0 \\ 0 \end{bmatrix}, \boldsymbol{x}_3 = \begin{bmatrix} 0 \\ 0 \\ 1 \\ 0 \end{bmatrix}, \boldsymbol{x}_4 = \begin{bmatrix} 0 \\ 0 \\ 0 \\ 1 \end{bmatrix}.$$

Thus for a given $d$, $\boldsymbol{x}_d \boldsymbol{x}_d^T$ has only one nonzero entry, specifically the $d$th diagonal entry. This implies that $\boldsymbol{B}$ can be treated as a diagonal matrix whose $d$th diagonal entry is $V_d$. Therefore, $\boldsymbol{B}^{-1}$ is a diagonal matrix whose $d$th diagonal entry is $V_d^{-1}$. Of course, this assumes that $\mu^{(d)}$ is neither 0 or 1.

Letting the notation $^{\otimes 2}$ represent the outer product of a vector with itself, the "meat" of the sandwich (the empirical covariance of the estimating function) is

$$\boldsymbol{M} = E\left(\left(\textstyle\sum_d w^{(d)} V_d^{-1} \boldsymbol{D}_d^T (Y - \mu^{(d)})\right)^{\otimes 2}\right).$$

In this context $\boldsymbol{D}_d^T = V_d \boldsymbol{x}_d^T$, so

$$\boldsymbol{M} = E\left(\left(\textstyle\sum_d w^{(d)} (Y - \mu^{(d)}) \boldsymbol{x}_d^T\right)^{\otimes 2}\right) = E\left(\begin{bmatrix} w^{(1)}(Y - \mu^{(1)}) \\ w^{(2)}(Y - \mu^{(2)}) \\ w^{(3)}(Y - \mu^{(3)}) \\ w^{(4)}(Y - \mu^{(4)}) \end{bmatrix}\begin{bmatrix} w^{(1)}(Y - \mu^{(1)}) \\ w^{(2)}(Y - \mu^{(2)}) \\ w^{(3)}(Y - \mu^{(3)}) \\ w^{(4)}(Y - \mu^{(4)}) \end{bmatrix}^T\right).$$

Therefore, a typical entry of $\boldsymbol{M}$ is

$$M_{dd'} = E\big(w^{(d)}w^{(d')}(Y - \mu^{(d)})(Y - \mu^{(d')})\big).$$

We can substitute the potential outcomes for the observed outcomes above by assumption 2, because only participants consistent with adaptive intervention $d$ will have nonzero $w^{(d)}$. Thus

$$M_{dd'} = E\big(w^{(d)}w^{(d')}(Y^{(d)} - \mu^{(d)})(Y^{(d')} - \mu^{(d')})\big).$$

To simplify the expression for $\boldsymbol{M_{dd'}}$ further, there are three cases to consider.

- **Case One**: If $d = d'$ then the entry is

$$M_{dd} = E\big((w^{(d)})^2(Y^{(d)} - \mu^{(d)})^2\big)$$

Using the law of iterated expectation,

$$M_{dd} = (1 - r_d)E\big((w^{(d)})^2(Y^{(d)} - \mu^{(d)})^2 | R^{(d)} = 0\big) +$$

$$r_d E\big((w^{(d)})^2(Y^{(d)} - \mu^{(d)})^2 | R^{(d)} = 1\big)$$

where $r_d$ is the response proportion for adaptive intervention $d$. As mentioned before, by the definition of the weights, nonresponders have a weight of 2 and are consistent with 1/2 of the embedded adaptive interventions, while responders have a weight of 4 and are consistent with 1/4 of the adaptive interventions. Therefore, by the consistency assumptions,

$$M_{dd} = (1 - r_d)E\big((4)^2 \tfrac{1}{4}(Y^{(d)} - \mu^{(d)})^2 | R^{(d)} = 0\big)$$

$$+ r_d E\big((2)^2 \tfrac{1}{2}(Y^{(d)} - \mu^{(d)})^2 | R^{(d)} = 1\big) = 4(1 - r_d)V_{d0} + 2r_dV_{d1},$$

where

$$V_{dr} = E\big((Y^{(d)} - \mu^{(d)})^2 | R^{(d)} = r\big).$$

In this context we use $r$ to denote a particular value 0 or 1 of the random variable $R^{(d)}$, although we caution that we follow other literature elsewhere in the paper in also using the symbol $r$ to indicate $P(R = 1)$ in cases where all $r_d$ are equal.

• **Case Two**: If $d \neq d'$ but the two adaptive interventions recommend the same first-stage intervention option, then $w^{(d)}w^{(d')}$ is nonzero for responders and zero for nonresponders. (Note that in this case that the probabilities $r_d$ and $r_{d'}$ are equal; response probability is assumed to depend only on first intervention option, as second assignment would not even occur in responders.) Therefore, by assumption 2 (consistency),

$$M_{dd} = 2r_d E\big((Y^{(d)} - \mu^{(d)})(Y^{(d')} - \mu^{(d')})|R^{(d)} = 1\big)$$

It is difficult to evaluate or elicit the cross-world covariance above, so we make the simplifying assumption 3 which says that it can be treated as zero because of randomization.

• **Case Three**: If $d \neq d'$ and the adaptive interventions differ on first-stage recommendations, then $w^{(d)}w^{(d')} = 0$ in all cases so that the cross-product can safely be ignored.

Therefore, $\boldsymbol{M}$ can be treated as a diagonal matrix with entries $4(1 - r_d)V_{d0} + 2r_d V_{d1}$.

The quantities $V_{d0}$ and $V_{d1}$ are not the same as the conditional variances for nonresponders and responders to adaptive intervention $d$, although they could be considered adjusted forms of the conditional variances, which add a positive term as described below.

Let $\mu^{(d)} = E(Y^{(d)})$ and $\psi^{(dr)} = E(Y^{(d)}|R = r)$. Then

$$V_{dr} = E((Y^{(d)} - \mu^{(d)})^2|R^{(d)} = r)$$

$$= E\left(\left((Y^{(d)} - \psi^{(dr)}) + (\psi^{(dr)} - \mu^{(d)})\right)^2 | R^{(d)} = r\right)$$

$$= E\left((Y^{(d)} - \psi^{(dr)})^2 | R^{(d)} = r\right) + \left(\psi^{(dr)} - \mu^{(d)}\right)^2$$

$$= \mathrm{Var}(Y^{(d)} | R^{(d)} = r) + \left(\psi^{(dr)} - \mu^{(d)}\right)^2 \tag{9}$$

treating $\mu^{(d)}$ and $\psi^{(dr)}$ as constants. Thus, $V_{dr}$ is actually equal to the conditional variance only in the very restricted case where $\psi^{(dr)} = \mu^{(d)}$, that is, when $E(Y^{(d)} | R^{(d)} = 0) = E(Y^{(d)} | R^{(d)} = 1) = E(Y^{(d)})$. This also means that $V_{dr}$ may in some cases be greater than $1/4$, even though the actual variance of a binary variable is bounded above by $1/4$. The additive terms can be calculated from conditional probabilities because $\mu^{(d)} = (1 - r_d)\psi^{(d0)} + r_d\psi^{(d1)}$ by iterative expectation, so that

$$\left(\psi^{(d0)} - \mu^{(d)}\right)^2 = \left(\psi^{(d0)} - ((1 - r_d)\psi^{(d0)} + r_d\psi^{(d1)})\right)^2 = r_d^2\left(\psi^{(d0)} - \psi^{(d1)}\right)^2$$

and similarly

$$\left(\psi^{(d1)} - \mu^{(d)}\right)^2 = \left(\psi^{(d1)} - ((1 - r_d)\psi^{(d0)} + r_d\psi^{(d1)})\right)^2$$

$$= (1 - r_d)^2\left(\psi^{(d0)} - \psi^{(d1)}\right)^2.$$

In this sense, the second term in (9) can be interpreted as the extra variability contributed by the tendency of responders and nonresponders to have different average outcomes to a given adaptive intervention for non-randomized reasons, above and beyond the effect of the second intervention option. The expressions can be combined so that $\boldsymbol{B}^{-1}\boldsymbol{M}\boldsymbol{B}^{-1}$ is a diagonal matrix with entries $(4(1 - r_d)V_{d0} + 2r_dV_{d1})/V_d^2$. Recall that the variance parameter $\sigma_\Delta^2$ for the power formula is the per-subject variance contribution to the estimate of the effect of interest, here the log odds ratio $\Delta = \boldsymbol{c}^T\boldsymbol{\theta}$. It is therefore equal to

$$\sigma_\Delta^2 = \frac{1}{n} \text{Var}(\boldsymbol{c}^T \boldsymbol{\theta}) = \boldsymbol{c}^T \text{Var}\left(\frac{1}{n}\boldsymbol{\theta}\right) \boldsymbol{c} = \boldsymbol{c}^T (\boldsymbol{B}^{-1} \boldsymbol{M} \boldsymbol{B}^{-1}) \boldsymbol{c}$$

$$= \boldsymbol{c}^T \text{Var}(\boldsymbol{B}^{-1} \boldsymbol{M} \boldsymbol{B}^{-1}) \boldsymbol{c} = \sum_d c_d^2 \frac{4(1-r_d)V_{d0} + 2r_d V_{d1}}{V_d^2},$$

where $\boldsymbol{c}$ is the contrast vector of interest. For a pairwise comparison between adaptive interventions $d$ and $d'$, the researcher would set $c_d = +1$, $c_{d'} = -1$, and other entries of $\boldsymbol{c}$ to zero. Thus, for such a pairwise comparison, assuming the adaptive interventions differ at least on first-stage assignment,

$$\sigma_\Delta^2 = \frac{4(1-r_d)V_{d0} + 2r_d V_{d1}}{V_d^2} + \frac{4(1-r_{d'})V_{d'0} + 2r_{d'} V_{d'1}}{V_{d'}^2}.$$

Substituting this expression for $\sigma_\Delta^2$ into Equation (1) leads to Equation (2).

### Derivation of Expression (3)

By the variance assumptions, $\sigma_\Delta^2 \leq 2(2-r_d)V_d^{-1} + 2(2-r_{d'})V_{d'}^{-1}$. This is then substituted into Expression (1).

### Derivation of Expression (4)

Suppose that a baseline measurement or pretest is being considered, so there are two observations per subject. Suppose the measurement times are coded as 0 and 1. We still solve the estimating equations given in expression (1), but now we consider the adaptive-intervention-specific expected value as a vector $\boldsymbol{\mu}^{(d)} = [\mu_0^{(d)}, \mu_1^{(d)}]^T$ rather than a scalar $\mu^{(d)}$. By the randomization design, the baseline measurement does not depend on the adaptive intervention that will be assigned, i.e., $\mu_0^{(d)} = \mu_0$ regardless of $d$. Therefore, the mean vector can be written as $\boldsymbol{\mu}^{(d)} = [\mu_0, \mu_d]^T$ with the same $\mu_0 = E(Y_0)$ for each adaptive intervention and with $\mu_d = \mu_1^{(d)}$.

Let $\eta_j = \text{logit}\left(\mu^{(j)}\right)$ for $j = 0,1,2,3,4$; these logits correspond to log odds of the outcome, and they equal the logistic regression coefficients in a dummy-coded parameterization described below. Below we derive the variance-covariance matrix of the vector of log odds parameters $\boldsymbol{\eta} = [\eta_0, \eta_1, \eta_2, \eta_3, \eta_4]^T$ using a sandwich formula

$$\text{Cov}(\widehat{\boldsymbol{\eta}}) = \frac{1}{n} \boldsymbol{B}^{-1} \boldsymbol{M} \boldsymbol{B}^{-1}$$

where matrices $\boldsymbol{B}$ and $\boldsymbol{M}$ are defined in the following subsection.

The outcome of interest is still assumed to be a log odds ratio comparing expected end-of-study outcomes under two adaptive interventions differing at least in first intervention option, that is, $\text{logit}\left(\mu^{(d)}\right) - \text{logit}\left(\mu^{(d')}\right)$, although now it is assumed that they will be estimated using a longitudinal model which takes within-subject correlation into account. The variance of the log odds ratio can be derived indirectly by noticing that $\Delta$ is a linear combination $\boldsymbol{c}^T \boldsymbol{\eta} = \sum_{j=0}^{4} c_j \, \eta_j$, in which $c_j$ is a vector having $+1$ in the position corresponding to adaptive intervention $d$, having $-1$ for the position corresponding to adaptive intervention $d'$, and having $0$ everywhere else. For example, for comparing the "enhanced working memory training" to the "enhanced incentives alone" adaptive interventions (Table 1), $\boldsymbol{c} = [0, +1, 0, -1, 0]^T$. Therefore,

$$\sigma_\Delta^2 = \boldsymbol{c}^T \boldsymbol{B}^{-1} \boldsymbol{M} \boldsymbol{B}^{-1} \boldsymbol{c}.$$

leading to the sample size recommendation (4). However, expression (4) is not useful on its own as a sample size formula unless $\boldsymbol{c}^T \boldsymbol{B}^{-1} \boldsymbol{M} \boldsymbol{B}^{-1} \boldsymbol{c}$ can be expressed in terms of quantities that have intuitive scientific or practical meaning and hence are impractical to elicit from investigators seeking to design a SMART. Suppose also that the weighted estimating equations of expression (8) will be used.

SMART BINARY

The adaptive-intervention-specific variance matrix of the potential outcomes is

$$\boldsymbol{V}_d = \text{Var}(\boldsymbol{Y}^{(d)}) = E((\boldsymbol{Y}^{(d)} - \mu^{(d)})^2)$$

$$= \begin{bmatrix} \text{Var}(Y_0^{(d)}) & \text{Cov}(Y_0^{(d)}, Y_1^{(d)}) \\ \text{Cov}(Y_0^{(d)}, Y_1^{(d)}) & \text{Var}(Y_1^{(d)}) \end{bmatrix} = \boldsymbol{G}_d^{\frac{1}{2}} \boldsymbol{R}_d \boldsymbol{G}_d^{\frac{1}{2}}$$

where

$$\boldsymbol{G}_d = \begin{bmatrix} \text{Var}(Y_0^{(d)}) & 0 \\ 0 & \text{Var}(Y_1^{(d)}) \end{bmatrix}$$

and

$$\boldsymbol{R}_d = \begin{bmatrix} 1 & \rho_d \\ \rho_d & 1 \end{bmatrix}.$$

The correlation parameter here is $\rho_d = \text{Corr}\left(Y_0^{(d)}, Y_1^{(d)}\right)$, and the Jacobian matrix is

$$\boldsymbol{D}_d = \boldsymbol{G}_d \boldsymbol{X}_d.$$

Therefore, Equation (1) becomes

$$\sum_{i=1}^n \sum_d w_i^{(d)} \boldsymbol{X}_d^T \boldsymbol{G}_d^{1/2} \boldsymbol{R}_d^{-1} \boldsymbol{G}_d^{-1/2} \left(\boldsymbol{Y}_i - \boldsymbol{\mu}^{(d)}(\boldsymbol{\theta})\right) = \boldsymbol{0}.$$

Several parameterizations are possible for constructing the $\boldsymbol{X}$ matrix. In practice, it may be good to use effect coding (see Dziak and colleagues, 2020). However, for purposes of the current derivation, a simpler parameterization is more convenient. Therefore the simplest parameterization would have a parameter for the general pretest expected value and a parameter for the posttest expected value of each adaptive intervention. That is, $E(Y_0^{(d)}) = \text{logit}^{-1}(\theta_0)$ and $E(Y_1^{(d)}) =$

logit$^{-1}(\theta_d)$. The $\theta$ parameters in this parameterization are the log odds corresponding to the probabilities given by the $\mu$ parameters.

In this parameterization, $\theta_0$ is not exactly an intercept, because it is not found in the $E(Y_1^{(d)}) = \theta_d$ terms. However, this parameterization expresses the same substantive assumptions as if we had assumed that $E(Y_0^{(d)}) = \theta_0$ and $E(Y_1^{(d)}) = \theta_0 + \theta_d^\star$, in which case $\theta_d^\star = \theta_d - \theta_0$ would represent an intervention effect rather than a mean. The contrasts of interest on the posttest will be the same in either parameterization, because $\theta_d^\star - \theta_{d'}^\star = (\theta_d - \theta_0) - (\theta_{d'} - \theta_0) = \theta_d - \theta_{d'}$. Our proposed parameterization is convenient because no parameter is shared between the expressions for the expected end-of-study outcome for any two embedded adaptive interventions. Specifically,

$$\boldsymbol{x}_1 = \begin{bmatrix} 1 & 0 \\ 0 & 1 \\ 0 & 0 \\ 0 & 0 \\ 0 & 0 \end{bmatrix}, \boldsymbol{x}_2 = \begin{bmatrix} 1 & 0 \\ 0 & 0 \\ 0 & 1 \\ 0 & 0 \\ 0 & 0 \end{bmatrix}, \boldsymbol{x}_3 = \begin{bmatrix} 1 & 0 \\ 0 & 0 \\ 0 & 0 \\ 0 & 1 \\ 0 & 0 \end{bmatrix}, \boldsymbol{x}_4 = \begin{bmatrix} 1 & 0 \\ 0 & 0 \\ 0 & 0 \\ 0 & 0 \\ 0 & 1 \end{bmatrix}.$$

Following Lu and colleagues (2016) and Seewald and colleagues (2020), we assume the asymptotic sandwich formula

$$\text{Cov}(\widehat{\boldsymbol{\theta}}) = \frac{1}{n}\boldsymbol{B}^{-1}\boldsymbol{M}\boldsymbol{B}^{-1}$$

but with a different $\boldsymbol{B}$ and $\boldsymbol{M}$ than were used in Derivation 2.

Let $v_0 = \mu^{(0)}(1 - \mu^{(0)})$ be the variance of the outcome at baseline (before the initial randomization), assumed to be the same for each adaptive intervention. Let $v_d = \mu^{(d)}(1 - \mu^{(d)})$ be the variance of the outcome at the final time point (end-of-study) for adaptive intervention $d$.

SMART BINARY

Let $\rho$ be the correlation between baseline outcome and the outcome at the final time point, assumed for simplicity to be the same for each embedded adaptive intervention.

Analogously to section 4.1, we make assumptions in order to approximate $\boldsymbol{B}$ and $\boldsymbol{M}$. The first two are consistency assumptions needed for identifiability, while the third and fourth are working assumptions used to provide a more tractable simplified formula for $\sigma_\Delta^2$.

1. **Consistency assumption** for $R$: same as assumption 1 in section 4.1.

2. **Consistency assumption** for $Y$: The observed **posttest** (end of study) outcome $Y_1$ for individuals consistent with adaptive intervention $d$ equals the potential outcome $Y_1^{(d)}$, which can be written $Y^{(d)}$ for short. It is a binary variable with expected value $\mu^{(d)}$ and variance $\mu^{(d)}(1 - \mu^{(d)})$. The observed **pretest** (baseline value of the outcome variable, before the intervention) $Y_0^{(d)}$ for a given individual is a constant regardless of the adaptive intervention which that individual will receive and can therefore be written as $Y^{(0)}$. Its expected value is $\mu^{(0)}$ and its variance is $\mu^{(0)}(1 - \mu^{(0)})$. The observed marginal pretest–posttest correlation is a constant $\rho$ for each adaptive intervention, that is, $\text{Corr}(Y_0^{(d)}, Y_1^{(d)}) = \rho$.

3. **Within-subject correlation:** Pretest and posttest residuals for each adaptive intervention are correlated within person at some nonnegative value, with

$$E\left( (Y_0^{(d)} - \mu_0^{(d)})(Y_1^{(d)} - \mu_1^{(d)}) \right) \equiv \rho_d \geq 0.$$

The marginal correlation $\rho_d$ is here assumed to be the same value $\rho$ for each adaptive intervention $d$ for purposes of deriving the formulas. Note that the correlation conditional on a value of $R$ need not be the same as the marginal correlation but is also assumed nonnegative:

$$E\left((Y_0^{(d)} - \mu_0^{(d)})(Y_1^{(d)} - \mu_1^{(d)})|R = r\right) \equiv \rho_d^r \geq 0.$$

4. **Cross-world independence assumption for responders**: For two adaptive interventions $d$ and $d'$ with the same initial intervention option $a_1$ but different $a_2$, cross`products of posttest residuals are independent:

$$E\left((\boldsymbol{Y}^{(d)} - \boldsymbol{\mu}^{(d)})(\boldsymbol{Y}^{(d')} - \boldsymbol{\mu}^{(d')})^T\right) = \boldsymbol{0}. \tag{10}$$

This last working assumption is somewhat unrealistic, but relaxing it would make the sample size formula much more complicated. Further, simulations presented in the main paper show that the simpler sample size formulas obtained by using this assumption perform very well in a fairly realistic scenario.

Generalizing the single-time-point results,

$$\boldsymbol{B} = E\left(\sum_d w^{(d)} \boldsymbol{D}_d^T \boldsymbol{V}_d^{-1} \boldsymbol{D}_d\right) = \sum_d E(w^{(d)}) \boldsymbol{D}_d^T \boldsymbol{V}_d^{-1} \boldsymbol{D}_d$$

$$= \sum_d \boldsymbol{D}_d^T \boldsymbol{V}_d^{-1} \boldsymbol{D}_d = \sum_d \boldsymbol{X}_d^T \boldsymbol{S}_d \boldsymbol{X}_d$$

where $\boldsymbol{S}_d = \boldsymbol{G}_d^{\frac{1}{2}} \boldsymbol{R}_d^{-1} \boldsymbol{G}_d^{\frac{1}{2}}$. Because of the special structure of $\boldsymbol{X}_1$,

$$\boldsymbol{X}_1^T \boldsymbol{S}_1 \boldsymbol{X}_1 = \begin{bmatrix} \boldsymbol{I}_{2\times 2} \\ \boldsymbol{0}_{3\times 2} \end{bmatrix} \boldsymbol{S}_1 [\boldsymbol{I}_{2\times 2} \quad \boldsymbol{0}_{3\times 2}] = \begin{bmatrix} \boldsymbol{S}_1 & \boldsymbol{0}_{2\times 3} \\ \boldsymbol{0}_{3\times 2} & \boldsymbol{0}_{3\times 3} \end{bmatrix} = \begin{bmatrix} (\boldsymbol{S}_1)_{11} & (\boldsymbol{S}_1)_{12} & 0 & 0 & 0 \\ (\boldsymbol{S}_1)_{21} & (\boldsymbol{S}_1)_{22} & 0 & 0 & 0 \\ 0 & 0 & 0 & 0 & 0 \\ 0 & 0 & 0 & 0 & 0 \\ 0 & 0 & 0 & 0 & 0 \end{bmatrix}$$

where $\boldsymbol{I}$ and $\boldsymbol{0}$ represent the identity and zero matrices of appropriate dimension.

SMART BINARY

Notice that, for example, $X_2$ has an analogous structure to $X_1$, but with the zeroes placed differently in the matrix. Therefore,

$$X_2^T S_2 X_2 = \begin{bmatrix} (S_2)_{11} & 0 & (S_2)_{12} & 0 & 0 \\ 0 & 0 & 0 & 0 & 0 \\ (S_2)_{21} & 0 & (S_2)_{22} & 0 & 0 \\ 0 & 0 & 0 & 0 & 0 \\ 0 & 0 & 0 & 0 & 0 \end{bmatrix}$$

Continuing in this way,

$$\sum_d X_d^T S_d X_d = \begin{bmatrix} \sum_d (S_d)_{11} & (S_1)_{12} & (S_2)_{12} & (S_3)_{12} & (S_4)_{12} \\ (S_1)_{21} & (S_1)_{22} & 0 & 0 & 0 \\ (S_2)_{21} & 0 & (S_2)_{22} & 0 & 0 \\ (S_3)_{21} & 0 & 0 & (S_3)_{22} & 0 \\ (S_4)_{21} & 0 & 0 & 0 & (S_4)_{12} \end{bmatrix}$$

To proceed from here we need an expression for $S_d$. Recall the parameterization which provides that $\mu_0^{(d)} = \beta_0$ and $\mu_1^{(d)} = \beta_d$. Therefore

$$S_d = G_d^{\frac{1}{2}} R_d^{-1} G_d^{\frac{1}{2}}$$

$$= \begin{bmatrix} \theta_0(1-\theta_0) & 0 \\ 0 & \theta_d(1-\theta_d) \end{bmatrix}^{1/2} \begin{bmatrix} 1 & \rho_d \\ \rho_d & 1 \end{bmatrix}^{-1} \begin{bmatrix} \theta_0(1-\theta_0) & 0 \\ 0 & \theta_d(1-\theta_d) \end{bmatrix}^{1/2}$$

$$= \frac{1}{1-\rho_d^2} \begin{bmatrix} v_0 & -\rho_d v_0^{1/2} v_d^{1/2} \\ -\rho_d v_0^{1/2} v_d^{1/2} & v_d \end{bmatrix}$$

where $v_0 = \theta_0(1-\theta_0) = \text{Var}(Y_0)$, which does not depend on $d$, and where $v_d = \theta_d(1-\theta_d) = \text{Var}(Y^{(d)})$.

Therefore, under the further simplifying assumption that $\rho_1 = \rho_2 = \ldots = \rho$,

$$\boldsymbol{B} = \sum_d \boldsymbol{X}_d^T \boldsymbol{S}_d \boldsymbol{X}_d =$$

$$\frac{1}{1-\rho^2}\begin{bmatrix} 4v_0 & -\rho v_0^{1/2}v_1^{1/2} & -\rho v_0^{1/2}v_2^{1/2} & -\rho v_0^{1/2}v_3^{1/2} & -\rho v_0^{1/2}v_4^{1/2} \\ -\rho v_0^{1/2}v_1^{1/2} & v_1 & 0 & 0 & 0 \\ -\rho v_0^{1/2}v_2^{1/2} & 0 & v_2 & 0 & 0 \\ -\rho v_0^{1/2}v_3^{1/2} & 0 & 0 & v_3 & 0 \\ -\rho v_0^{1/2}v_4^{1/2} & 0 & 0 & 0 & v_4 \end{bmatrix}.$$

The bread matrix $\boldsymbol{B}^{-1}$ is the inverse of the matrix above. Using the formula of Salkuyeh and Beik (2018),

$$\boldsymbol{B}^{-1} = \begin{bmatrix} \frac{1}{4v_0} & \frac{\rho}{4\sqrt{v_0 v_1}} & \frac{\rho}{4\sqrt{v_0 v_2}} & \frac{\rho}{4\sqrt{v_0 v_3}} & \frac{\rho}{4\sqrt{v_0 v_4}} \\ \frac{\rho}{4\sqrt{v_0 v_1}} & \frac{4-3\rho^2}{4v_1} & \frac{\rho^2}{4\sqrt{v_1 v_2}} & \frac{\rho^2}{4\sqrt{v_1 v_3}} & \frac{\rho^2}{4\sqrt{v_1 v_4}} \\ \frac{\rho}{4\sqrt{v_0 v_2}} & \frac{\rho^2}{4\sqrt{v_1 v_2}} & \frac{4-3\rho^2}{4v_2} & \frac{\rho^2}{4\sqrt{v_2 v_3}} & \frac{\rho^2}{4\sqrt{v_2 v_4}} \\ \frac{\rho}{4\sqrt{v_0 v_3}} & \frac{\rho^2}{4\sqrt{v_1 v_3}} & \frac{\rho^2}{4\sqrt{v_2 v_3}} & \frac{4-3\rho^2}{4v_3} & \frac{\rho^2}{4\sqrt{v_3 v_4}} \\ \frac{\rho}{4\sqrt{v_0 v_4}} & \frac{\rho^2}{4\sqrt{v_1 v_4}} & \frac{\rho^2}{4\sqrt{v_2 v_4}} & \frac{\rho^2}{4\sqrt{v_3 v_4}} & \frac{4-3\rho^2}{4v_4} \end{bmatrix}$$

Next, let $\boldsymbol{V}_{dr}$ be the diagonal matrix with diagonal entries $V_{0r}$ and $V_{dr}$ of adjusted conditional variances of the baseline measurement and of the final measurement given adaptive intervention $d$ and response status $R$. The adjusted posttest conditional variances $V_{d0} = E\left((Y_0 - \mu^{(0)})^2 | R = 0\right)$ and $V_{d1} = E\left((Y_0 - \mu^{(0)})^2 | R = 1\right)$ are the same as in the one-time, posttest-only case described earlier, and the pretest conditional variance $V_{00}$ and $V_{01}$ can similarly be calculated from elicited probabilities as

$$V_{00} = \psi^{(00)}\left(1 - \psi^{(00)}\right) + \left(\psi^{(d0)} - \mu^{(d)}\right)^2$$
$$V_{01} = \psi^{(01)}\left(1 - \psi^{(01)}\right) + \left(\psi^{(d1)} - \mu^{(d)}\right)^2,$$

letting $\psi^{(00)} = E(Y_0|R = 0)$ and $\psi^{(01)} = E(Y_0|R = 1)$. In particular, under the unrealistic assumption that future responder status is not related to baseline outcome, $V_{00}$ and $V_{01}$ would each simplify to $v_0$; otherwise they will be somewhat larger due to the added positive term.

Then, the "meat" of the sandwich is

$$M = E\left(\left(\sum_d w^{(d)} \boldsymbol{D}_d^T \boldsymbol{V}_d^{-1}(\boldsymbol{Y} - \boldsymbol{\mu}^{(d)})\right)^{\otimes 2}\right).$$

Following Seewald and colleagues (2020), we expand $\boldsymbol{M}$ as

$$\boldsymbol{M} = \sum_d E\left(\left(w^{(d)} \boldsymbol{D}_d^T \boldsymbol{V}_d^{-1}(\boldsymbol{Y} - \boldsymbol{\mu}^{(d)})\right)^{\otimes 2}\right) +$$

$$\sum_d \sum_{d' \neq d} E\left(w^{(d)} w^{(d')} \left(\boldsymbol{D}_d^T \boldsymbol{V}_d^{-1}(\boldsymbol{Y} - \boldsymbol{\mu}^{(d)})\right)\left(\boldsymbol{D}_{d'}^T \boldsymbol{V}_{d'}^{-1}(\boldsymbol{Y} - \boldsymbol{\mu}^{(d')})\right)^T\right). \quad (11)$$

Consider terms in the first summation. Using iterated expectation,

$$E\left(\left(w^{(d)} \boldsymbol{D}_d^T \boldsymbol{V}_d^{-1}(\boldsymbol{Y} - \boldsymbol{\mu}^{(d)})\right)^{\otimes 2}\right) =$$

$$(1 - r_d)E\left(\left(w^{(d)} \boldsymbol{D}_d^T \boldsymbol{V}_d^{-1}(\boldsymbol{Y} - \boldsymbol{\mu}^{(d)})\right)^{\otimes 2} | R^{(d)} = 0\right) +$$

$$r_d E\left(\left(w^{(d)} \boldsymbol{D}_d^T \boldsymbol{V}_d^{-1}(\boldsymbol{Y} - \boldsymbol{\mu}^{(d)})\right)^{\otimes 2} | R^{(d)} = 1\right)$$

$$= (1 - r_d)E\left((w^{(d)})^2 \boldsymbol{D}_d^T \boldsymbol{V}_d^{-1}(\boldsymbol{Y} - \boldsymbol{\mu}^{(d)})^{\otimes 2} \boldsymbol{V}_d^{-1} \boldsymbol{D}_d | R^{(d)} = 0\right) +$$

$$r_d E\left((w^{(d)})^2 \boldsymbol{D}_d^T \boldsymbol{V}_d^{-1}(\boldsymbol{Y} - \boldsymbol{\mu}^{(d)})^{\otimes 2} \boldsymbol{V}_d^{-1} \boldsymbol{D}_d | R^{(d)} = 1\right).$$

Using the consistency assumptions, and the definition of the weights $w_i^{(d)}$ which give zero weight to participants not consistent with a given adaptive intervention $d$, we can replace $\boldsymbol{Y}$ above with $\boldsymbol{Y}^{(d)}$. Also,

SMART BINARY

$$E\left((w^{(d)})^2 \boldsymbol{D}_d^T \boldsymbol{V}_d^{-1} (\boldsymbol{Y}^{(d)} - \boldsymbol{\mu}^{(d)})^{\otimes 2} \boldsymbol{V}_d^{-1} \boldsymbol{D}_d | R^{(d)} = 1\right) = 2\boldsymbol{D}_d^T \boldsymbol{V}_d^{-1} \boldsymbol{V}_{d1} \boldsymbol{V}_d^{-1} \boldsymbol{D}_d$$

where $\boldsymbol{V}_{d0} = E\left((\boldsymbol{Y}^{(d)} - \boldsymbol{\mu}^{(d)})^{\otimes\;2} | R = 0\right)$ and $\boldsymbol{V}_{d1} = E\left((\boldsymbol{Y}^{(d)} - \boldsymbol{\mu}^{(d)})^{\otimes\;2} | R = 1\right)$. Therefore,

$$E\left(\left(w^{(d)} \boldsymbol{D}_d^T \boldsymbol{V}_d^{-1} (\boldsymbol{Y} - \boldsymbol{\mu}^{(d)})\right)^{\otimes 2}\right) = 4(1 - r_d) \boldsymbol{D}_d^T \boldsymbol{V}_d^{-1} \boldsymbol{V}_{d0} \boldsymbol{V}_d^{-1} \boldsymbol{D}_d + 2r_d \boldsymbol{D}_d^T \boldsymbol{V}_d^{-1} \boldsymbol{V}_{d1} \boldsymbol{V}_d^{-1} \boldsymbol{D}_d.$$

Notice that $\boldsymbol{V}_{dr}$ is not exactly the same as $\text{Cov}(\boldsymbol{Y}^{(d)} | R^{(d)} = r)$ because the residual vector being squared is not $\boldsymbol{Y}^{(d)} - E(\boldsymbol{Y}^{(d)} | R^{(d)} = r)$ but instead $\boldsymbol{Y}^{(d)} - E(\boldsymbol{Y}^{(d)})$. However, $\boldsymbol{V}_{dr}$ can be calculated indirectly by considering that $E(Y^{(d)} - E(Y^{(d)}|R))^2 = E(Y^{(d)} - E(Y^{(d)}))^2 + (E(Y^{(d)}|R) - E(Y^{(d)}))^2$. Thus $\boldsymbol{V}_{dr}$ can be calculated using the conditional probabilities $E(Y^{(d)}|R)$ if those are elicited.

Next consider an off-diagonal term in (11), specifically

$$E\left(w^{(d)} w^{(d')} \left(\boldsymbol{D}_d^T \boldsymbol{V}_d^{-1} (\boldsymbol{Y} - \boldsymbol{\mu}^{(d)})\right) \left(\boldsymbol{D}_{d'}^T \boldsymbol{V}_{d'}^{-1} (\boldsymbol{Y} - \boldsymbol{\mu}^{(d')})\right)^T\right)$$

If adaptive interventions $d$ and $d'$ do not recommend the same initial intervention option, then no participant can be compatible with both, and this cross-product is zero because $w^{(d)} w^{(d')}$ is zero. However, if adaptive interventions $d$ and $d'$ recommend the same initial intervention option, then $w^{(d)} w^{(d')}$ can be nonzero for responders. In this case, using iterated expectation, the off-diagonal cross-product equals

$$E\left(w^{(d)} w^{(d')} \left(\boldsymbol{D}_d^T \boldsymbol{V}_d^{-1} (\boldsymbol{Y} - \boldsymbol{\mu}^{(d)})\right) \left(\boldsymbol{D}_{d'}^T \boldsymbol{V}_{d'}^{-1} (\boldsymbol{Y} - \boldsymbol{\mu}^{(d')})\right)^T\right)$$

$$= 0 + r_d E\left(w^{(d)} w^{(d')} \left(\boldsymbol{D}_d^T \boldsymbol{V}_d^{-1} (\boldsymbol{Y} - \boldsymbol{\mu}^{(d)})\right) \left(\boldsymbol{D}_{d'}^T \boldsymbol{V}_{d'}^{-1} (\boldsymbol{Y} - \boldsymbol{\mu}^{(d')})\right)^T | R = 1\right)$$

$$= r_d \boldsymbol{D}_d^T \boldsymbol{V}_d^{-1} \boldsymbol{V}_{dd'1} \boldsymbol{V}_{d'}^{-1} \boldsymbol{D}_{d'}$$

where $\boldsymbol{V}_{dd'1} = E\big((\boldsymbol{Y}^{(d)} - \boldsymbol{\mu}^{(d)})(\boldsymbol{Y}^{(d')} - \boldsymbol{\mu}^{(d')})^T | R = 1\big)$. This $\boldsymbol{V}_{dd'1}$ is not necessarily zero in practice, but it is very hard to calculate or to interpret, and so we propose to use expression (10) in assumption 4 to assume that it is zero for purposes of deriving the power formula.

Under this assumption, the extra terms in (11) disappear so that

$$\boldsymbol{M} = \sum_d 4(1 - r_d)\boldsymbol{D}_d^T\boldsymbol{V}_d^{-1}\boldsymbol{V}_{d0}\boldsymbol{V}_d^{-1}\boldsymbol{D}_d + \sum_d 2r_d\boldsymbol{D}_d^T\boldsymbol{V}_d^{-1}\boldsymbol{V}_{d1}\boldsymbol{V}_d^{-1}\boldsymbol{D}_d. \qquad (12)$$

Then

$$\boldsymbol{M} = \sum_d (\boldsymbol{X}_d^T Q_d \boldsymbol{X}_d) = \begin{bmatrix} \sum_d Q_{11}^{(d)} & Q_{21}^{(1)} & Q_{21}^{(2)} & Q_{21}^{(3)} & Q_{21}^{(4)} \\ Q_{12}^{(1)} & Q_{22}^{(1)} & 0 & 0 & 0 \\ Q_{12}^{(2)} & 0 & Q_{22}^{(2)} & 0 & 0 \\ Q_{12}^{(3)} & 0 & 0 & Q_{22}^{(3)} & 0 \\ Q_{12}^{(4)} & 0 & 0 & 0 & Q_{22}^{(4)} \end{bmatrix}$$

where $\boldsymbol{G}_d = \begin{bmatrix} v_0 & 0 \\ 0 & v_d \end{bmatrix}$ and $\boldsymbol{R}_d = \begin{bmatrix} 1 & \rho \\ \rho & 1 \end{bmatrix}$, and

$$\boldsymbol{Q}^{(d)} = \begin{bmatrix} Q_{11}^{(d)} & Q_{12}^{(d)} \\ Q_{21}^{(d)} & Q_{22}^{(d)} \end{bmatrix} = \boldsymbol{G}_d^{1/2}\boldsymbol{R}_d^{-1}\boldsymbol{G}_d^{-1/2}(4(1 - r_d)\boldsymbol{V}_{d0} + 2r_d\boldsymbol{V}_{d1})\boldsymbol{G}_d^{1/2}\boldsymbol{R}_d^{-1}\boldsymbol{G}_d^{-1/2}.$$

We now have expressions, although not very simple ones, for $\boldsymbol{B}^{-1}$ and $\boldsymbol{M}$. One could therefore substitute $\sigma_\Delta^2 = \boldsymbol{B}^{-1}\boldsymbol{M}\boldsymbol{B}^{-1}$ in to equations (6). However, this approach could be challenging in practice. This is because for each adaptive intervention $d$, it requires values for $\boldsymbol{V}_{d0}$ and $\boldsymbol{V}_{d1}$ in addition to the marginal covariance matrix $\boldsymbol{V}_d$. The quantities $\boldsymbol{V}_{d0}$ and $\boldsymbol{V}_{d1}$ do not have an intuitive interpretation themselves as quantities of scientific interest.

This issue is not insurmountable, because $\boldsymbol{V}_{d0}$ and $\boldsymbol{V}_{d1}$ can be computed indirectly from the $\psi$ and $\mu$ parameters, while the $\mu$ parameters can be computed from the $\psi$ parameters and response rates $r_d$. Lastly, the $\psi$ parameters can be elicited indirectly as conditional probabilities for outcomes in appropriate cells. For example, for adaptive intervention $(+1, +1)$, the value of $\psi_{d0}$ should equal the expected outcome probability in the observed cell with $a_1 = +1$, $R = 0$, and $a_2 = +1$. Similarly, $\psi_{d1}$ would equal the expected outcome probability in the observed cell with $a_1 = +1$, $R = 1$, and $a_2 = +1$. There are only six observed cells, despite eight possible combinations of $d = 1,2,3,4$ and $r = 0,1$, but this is reasonable because $\psi_{d1}$ will equal $\psi_{d^\star 1}$ for a pair of adaptive interventions $d$ and $d^\star$ differing only on $a_1$. That is, under reasonable elicitation assumptions, $P(Y(a_1, a_{2NR}) = 1 | R(a_1, a_{2NR} = 1)$ should not depend on $a_{2NR}$. That is, the hypothetical choice of second intervention option $a_2$ does not affect the outcome for responders, who never receive this part of the adaptive intervention. Nonetheless, even though $\boldsymbol{M}$ can be calculated using expression (12), a simpler formula would be desirable.

**Derivation of Expression (5)**

If one is willing to assume $\boldsymbol{V}_{d0} \approx \boldsymbol{V}_{d1} \approx \boldsymbol{V}_d$ for each $d$, i.e., that the variance is independent of response status, and also that the response rate $r_d$ is the same for each $d$, then $\sigma_\Delta^2 = \boldsymbol{c}^T \boldsymbol{B}^{-1} \boldsymbol{M} \boldsymbol{B}^{-1} \boldsymbol{c} \approx 2(2 - r) \boldsymbol{c}^T \boldsymbol{B}^{-1} \boldsymbol{c}$. Then $\boldsymbol{c}^T \boldsymbol{B}^{-1} \boldsymbol{c}$ can be shown to equal

$$\frac{4 - 3\rho^2}{4v_d} - \frac{\rho^2}{2\sqrt{v_d v_{d'}}} + \frac{4 - 3\rho^2}{4v_{d'}}.$$

Combining this with expression (1) leads to the sample size recommendation.

**Alternative to Expression (5)**

It was noted earlier that the working independence assumption (10) is somewhat unrealistic. This is because it implies that

$$E\left((Y_0^{(d)} - \mu^{(d)})(Y_1^{(d')} - \mu^{(d')})|R^{(d)} = 1\right) = 0$$

and

$$E\left(Y_1^{(d)} - \mu^{(d)})(Y_0^{(d')} - \mu^{(d')})|R^{(d)} = 1\right) = 0.$$

This cross-world independence is standard after randomization, but less intuitive before randomization. It would be much more parsimonious to make the assumption of cross-world independence only for the final outcomes and not for the baseline outcomes, i.e., to assume

$$E\left((Y_1^{(d)} - \mu_0^{(d)})(Y_1^{(d')} - \mu_1^{(d')})|R^{(d)} = 1\right) = 0 \tag{13}$$

instead of (10). Therefore, it is worth considering what the $2 \times 2$ matrix $\boldsymbol{V}_{dd'1}$ would equal if equation (13) was true but equation (10) was not true; that is, post-randomization potential outcomes are independent across counterfactual worlds but pre-randomization potential outcomess are identical among them. In this case, because posttest residuals are still independent, the lower right corner of $\boldsymbol{V}_{dd'1}$ is still zero. However, the off-diagonal entries of $\boldsymbol{V}_{dd'1}$ are not zero:

$$(\boldsymbol{V}_{dd'1})_{12} = E\left((Y_0^{(d)} - \mu_0^{(d)})(Y_1^{(d')} - \mu_1^{(d')})|R = 1\right)$$

$$= E\left((Y_0^{(d')} - \mu_0^{(d')})(Y_1^{(d')} - \mu_1^{(d')})|R = 1\right)$$

In this subsection we only need to consider pairs of adaptive interventions agreeing on first stage intervention option $a_1$, because other pairs will have weight zero in (11). Therefore, we can at least assume that $R^{(d)} = R^{(d')}$. Then

$$\boldsymbol{V}_{dd'1} = E\left((\boldsymbol{Y}^{(d)} - \boldsymbol{\mu}^{(d)})(\boldsymbol{Y}^{(d')} - \boldsymbol{\mu}^{(d')})^T)|R^{(d)} = 1\right)$$

SMART BINARY

$$= E\left(\left((\boldsymbol{Y}^{(d)} - \boldsymbol{\psi}^{(d,1)}) + (\boldsymbol{\psi}^{(d,1)} - \boldsymbol{\mu}^{(d)})\right)\left((\boldsymbol{Y}^{(d')} - \boldsymbol{\psi}^{(d',1)}) + (\boldsymbol{\psi}^{(d',1)} - \boldsymbol{\mu}^{(d')})\right)|R^{(d)} = 1\right)$$

$$= E\left((\boldsymbol{Y}^{(d')} - \boldsymbol{\psi}^{(d',1)})^2|R^{(d)} = 1\right) + E\left((\boldsymbol{Y}^{(d)} - \boldsymbol{\psi}^{(d,1)})|R^{(d)} = 1\right)\left(\boldsymbol{\psi}^{d',1} - \boldsymbol{\mu}^{d'}\right)^T$$

$$+ E\left((\boldsymbol{Y}^{(d')} - \boldsymbol{\psi}^{(d',1)})|R^{(d')} = 1\right)\left(\boldsymbol{\psi}^{(d,1)} - \boldsymbol{\mu}^{(d)}\right)^T + (\boldsymbol{\psi}^{(d,1)} - \boldsymbol{\mu}^{(d)})(\boldsymbol{\psi}^{(d',1)-\boldsymbol{\mu}^{(d')}})^T$$

$$= \begin{bmatrix} \text{Var}(Y^{(0)}|R = 1) & \text{Cov}(Y^{(0)}, Y^{(d)}|R = 1) \\ \text{Cov}(Y^{(0)}, Y^{(d')}|R = 1) & \text{Cov}(Y^{(d)}, Y^{(d')}|R = 1) \end{bmatrix}$$

$$+ \begin{bmatrix} (\psi^{(0,1)} - \mu^{(0)}) & (\psi^{(0,1)} - \mu^{(0)})(\psi^{(d,1)} - \mu^{(d)}) \\ (\psi^{(0,1)} - \mu^{(0)})(\psi^{(d',1)} - \mu^{(d')}) & (\psi^{(d,1)} - \mu^{(d)})(\psi^{(d',1)} - \mu^{(d')}) \end{bmatrix}$$

$$= \begin{bmatrix} \psi^{(0,1)}(1 - \psi^{(0,1)}) & \rho^\star\sqrt{\psi^{(0,1)}(1 - \psi^{(0,1)})\psi^{(d,1)}(1 - \psi^{(d,1)})} \\ \rho^\star\sqrt{\psi^{(0,1)}(1 - \psi^{(0,1)})\psi^{(d',1)}(1 - \psi^{(d',1)})} & 0 \end{bmatrix}$$

$$+ \begin{bmatrix} (\psi^{(0,1)} - \mu^{(0)})^2 & (\psi^{(0,1)} - \mu^{(0)})(\psi^{(d,1)} - \mu^{(d)}) \\ (\psi^{(0,1)} - \mu^{(0)})(\psi^{(d',1)} - \mu^{(d')}) & (\psi^{(d,1)} - \mu^{(d)})(\psi^{(d',1)} - \mu^{(d')}) \end{bmatrix}$$

where $\psi^{(0,1)} = E(Y_0|R = 1)$ and $\psi^{(d,1)} = E(Y_1^{(d)}|R = 1)$, and where $\rho^\star$ is the pretest-posttest correlation conditional on $R = 1$, assumed here to be the same for each adaptive intervention. However, this expression is likely to be more difficult for a substantive researcher to use. Fortunately, the exact error variance of the pretest mean (or intercept) does not matter very much to the contrasts of interest in this paper, which focuses on pairwise end-of-study contrasts between adaptive interventions differing in initial intervention option $a_1$. Accordingly, in our simulations the power formula still performs quite well, despite working assumption (10).

**Alternative Formulas with Identity Link Function (Linear Model)**

In this paper we have used an approach similar to logistic regression, with a logit link function. Other options would be to use a log link or an identity link. Although the identity link is generally not recommended for binary data, it is helpful to derive formulas for this case because this was the link used for linear modeling by Seewald and colleagues (2020), and it therefore allows the formulas to be compared directly with theirs. This would involve using linear mean function $\boldsymbol{\mu}^{(d)} = \boldsymbol{X\theta}$ in Equation (1) instead of the logit link function. Accordingly we would assume the estimand $\Delta$ of interest in Equation (1) would be a difference in probabilities (risk difference). Then the estimand $\Delta$ could still be written as a linear combination $\boldsymbol{c}^T\boldsymbol{\theta}$ of the new coefficients $\boldsymbol{\theta}$, using the same $\boldsymbol{c}$ coefficients as before. The power and sample size formulas (1) and (2) could still be used, but a new definition of $\sigma_\Delta^2$ and of $\mathrm{Cov}(\widehat{\boldsymbol{\theta}})$ would be needed. However, despite the new link function we still should not treat the binary $Y$ is homoskedastic. The variance $\mathrm{Var}(Y_t^{(d)})$ would depend on adaptive intervention and time.

In the cross-sectional (single-wave) case, with $Y$ as the end-of-study binary outcome, we still have the same form $V_d = \mu^{(d)}(1 - \mu^{(d)})$, because that is dictated by the distributional form of $Y$ and not by the link function or estimand of interest. However, the derivative $\boldsymbol{D}_d$ of the link function is now simply $\boldsymbol{x}_d$ instead of $(\mu^{(d)}(1 - \mu^{(d)}))^{-1}\boldsymbol{x}_d$. Therefore, the simplified form of Equation (8) is the weighted least squares equation

$$\sum_{i=1}^n \sum_d w_i^{(d)} V_d^{-1} \boldsymbol{x}_d^T (Y_i - \boldsymbol{x}_d \boldsymbol{\theta}) = \boldsymbol{0}.$$

The bread matrix is

$$\sum_d V_d^{-1} \boldsymbol{D}_d^T \boldsymbol{D}_d = \sum_d V_d^{-1} V_d^2 \boldsymbol{x}_d^T \boldsymbol{x}_d = \sum_d V_d \boldsymbol{x}_d^T \boldsymbol{x}_d,$$

SMART BINARY

which simplifies to a diagonal matrix whose $d$th diagonal entry is $V_d$, exactly the inverse of the bread matrix obtained before for the logistic link function. Thus $V_d$ is used here where $V_d^{-1}$ was used in (6). Intuitively, this discrepancy occurs because we are now studying probabilities (which are constrained above and below and therefore have more sampling variance for common events than rare ones) than odds (which are based on ratios and therefore have more sampling variance for rare events than common ones).

The meat matrix is now

$$\boldsymbol{M} = E\left(\left(\sum_d w^{(d)} V_d^{-1}(Y - \mu^{(d)}) \boldsymbol{x}_d^T\right)^{\otimes 2}\right),$$

because $V_d^{-1}$ no longer cancels out. Going through the same steps as in Section 4.3 would lead to $\boldsymbol{M}$ as a diagonal matrix with entries

$$M_{dd'} = E(w^{(d)} w^{(d')} V_d^{-2} (Y - \mu^{(d)})(Y - \mu^{(d')})$$

$$= \frac{4(1 - r_d)V_{d0} + 2r_d V_{d1}}{V_d^2}.$$

Therefore, $\boldsymbol{B}^{-1}\boldsymbol{M}\boldsymbol{B}^{-1}$ becomes a diagonal matrix whose entries are simply

$$4(1 - r_d)V_{d0} + 2r_d V_{d1}.$$

This means that

$$\sigma_\Delta^2 = \sum_d c_d^2 (4(1 - r_d)V_{d0} + 2r_d V_{d1})$$

and that for a contrast of two adaptive interventions $d$ and $d'$ differing at least on first stage assignment,

$$\sigma_\Delta^2 = 4(1 - r_d)V_{d0} + 2r_d V_{d1} + 4(1 - r_{d'})V_{d'0} + 2r_{d'} V_{d'1}.$$

SMART BINARY

Under the assumptions of Kidwell and colleagues (2019) that $V_{d0} \leq V_d$, $V_{d1} \leq V_d$, $V_{d'0} \leq V_{d'}$, $V_{d'1} \leq V_{d'}$, and $r_d = r_{d'} = r$, this leads to

$$\sigma_\Delta^2 \leq 4(1-r_d)V_d + 2r_d V_d + 4(1-r_{d'})V_{d'0} + 2r_{d'}V_{d'1}$$

$$= 2(2-r)(V_d + V_{d'}).$$

This leads to the sample size formula

$$n = \frac{2(2-r)\left(z_q + z_{1-\alpha/2}\right)^2 (V_d + V_{d'})}{\Delta^2} \tag{14}$$

so that the contributions of the variances are again the reciprocals of what they were in the logistic case. Notice that $V_d$ and $V_{d'}$ play very different roles in (14) as in Equation (3) due to the different estimands (probability versus log odds). However, this should not be taken to mean that power is necessarily higher for one test or the other, because the effect size $\Delta$ would also not be the same under the two modeling approaches.

Under the unrealistic further assumption that $V_d = V_{d'} = \sigma^2$ for some quantity $\sigma^2$, the above expression would become the usual sample size formula for a comparison of two means (e.g., pooled $z$ or approximate pooled $t$ test), multiplied by the usual design effect $2 - r$.

Now consider the linear link in the pretest–posttest case, with $E(Y_0^{(d)}) = \theta_0$ and $E(Y_1^{(d)}) = \theta_d$. In this case, the matrices $\boldsymbol{X}_d$, $\boldsymbol{G}_d$ and $\boldsymbol{R}_d$ are the same as in section (5). Also as before, the variance matrix $\boldsymbol{V}_d$ is

$$\boldsymbol{G}_d^{-\frac{1}{2}}\boldsymbol{R}_d^{-1}\boldsymbol{G}_d^{-\frac{1}{2}} = \frac{1}{1-\rho_d^2}\begin{bmatrix} v_0^{-1} & -\rho_d v_0^{-1/2}v_d^{-1/2} \\ -\rho_d v_0^{-1/2}v_d^{-1/2} & v_d^{-1}. \end{bmatrix}$$

SMART BINARY

However, $\boldsymbol{D}_d = \boldsymbol{X}_d$ instead of $\boldsymbol{G}_d\boldsymbol{X}_d$ as before. The bread matrix becomes $\sum_d \boldsymbol{D}_d^T\boldsymbol{V}_d^{-1}\boldsymbol{D}_d = \sum_d \boldsymbol{X}_d^T\boldsymbol{V}_d^{-1}\boldsymbol{X}_d$. Next,

$$\sum_d \boldsymbol{X}_d^T\boldsymbol{V}_d^{-1}\boldsymbol{X}_d = \begin{bmatrix} \sum_d (\boldsymbol{V}_d^{-1})_{11} & (\boldsymbol{V}_1^{-1})_{12} & (\boldsymbol{V}_2^{-1})_{12} & (\boldsymbol{V}_3^{-1})_{12} & (\boldsymbol{V}_4^{-1})_{12} \\ (\boldsymbol{V}_1^{-1})_{21} & (\boldsymbol{V}_1^{-1})_{22} & 0 & 0 & 0 \\ (\boldsymbol{V}_2^{-1})_{21} & 0 & (\boldsymbol{V}_2^{-1})_{22} & 0 & 0 \\ (\boldsymbol{V}_3^{-1})_{21} & 0 & 0 & (\boldsymbol{V}_3^{-1})_{22} & 0 \\ (\boldsymbol{V}_4^{-1})_{21} & 0 & 0 & 0 & (\boldsymbol{V}_4^{-1})_{12} \end{bmatrix}$$

Assuming that $\rho_1 = \rho_2 = \ldots = \rho$, $\boldsymbol{B}$ would be

$$\frac{1}{1-\rho^2}\begin{bmatrix} 4v_0^{-1} & -\rho v_0^{-1/2}v_1^{-1/2} & -\rho v_0^{-1/2}v_2^{-1/2} & -\rho v_0^{-1/2}v_3^{-1/2} & -\rho v_0^{-1/2}v_4^{-1/2} \\ -\rho v_0^{-1/2}v_1^{-1/2} & v_1^{-1} & 0 & 0 & 0 \\ -\rho v_0^{-1/2}v_2^{-1/2} & 0 & v_2^{-1} & 0 & 0 \\ -\rho v_0^{-1/2}v_3^{-1/2} & 0 & 0 & v_3^{-1} & 0 \\ -\rho v_0^{-1/2}v_4^{-1/2} & 0 & 0 & 0 & v_4^{-1} \end{bmatrix}$$

where once again, the most noticeable difference in form relative to the logistic case is that all variances are replaced by their reciprocals. Thus, using the Salkuyeh and Beik (2018) formula again,

$$\boldsymbol{B}^{-1} = \frac{1}{4}\begin{bmatrix} v_0 & \rho\sqrt{v_0 v_1} & \rho\sqrt{v_0 v_2} & \rho\sqrt{v_0 v_3} & \rho\sqrt{v_0 v_4} \\ \rho\sqrt{v_0 v_1} & (4-3\rho^2)v_1 & \rho^2\sqrt{v_1 v_2} & \rho^2\sqrt{v_1 v_3} & \rho^2\sqrt{v_1 v_4} \\ \rho\sqrt{v_0 v_2} & \rho^2\sqrt{v_1 v_2} & (4-3\rho^2)v_2 & \rho^2\sqrt{v_2 v_3} & \rho^2\sqrt{v_2 v_4} \\ \rho\sqrt{v_0 v_3} & \rho^2\sqrt{v_1 v_3} & \rho^2\sqrt{v_2 v_3} & (4-3\rho^2)v_3 & \rho^2\sqrt{v_3 v_4} \\ \rho\sqrt{v_0 v_4} & \rho^2\sqrt{v_1 v_4} & \rho^2\sqrt{v_2 v_4} & \rho^2\sqrt{v_3 v_4} & (4-3\rho^2)v_4 \end{bmatrix}.$$

The meat matrix would then be

$$\boldsymbol{M} = E\left(\left(\sum_d w^{(d)} \boldsymbol{X}_d^T \boldsymbol{V}_d^{-1}(\boldsymbol{Y} - \boldsymbol{x}_d \boldsymbol{\theta})\right)^{\otimes 2}\right).$$

The cross-product within adaptive intervention $d$ would be

$$E\left(\left(w^{(d)} \boldsymbol{X}_d^T \boldsymbol{V}_d^{-1}(\boldsymbol{Y} - \boldsymbol{\mu}^{(d)})\right)^{\otimes 2}\right) = 4(1 - r_d)\boldsymbol{X}_d^T \boldsymbol{V}_d^{-1} \boldsymbol{V}_{d0} \boldsymbol{V}_d^{-1} \boldsymbol{X}_d$$

$$+2r_d \boldsymbol{X}_d^T \boldsymbol{V}_d^{-1} \boldsymbol{V}_{d1} \boldsymbol{V}_d^{-1} \boldsymbol{X}_d.$$

If we assume $\boldsymbol{V}_{dr} \approx \boldsymbol{V}$ then the above approximates $2(2 - r_d)\boldsymbol{X}_d^T \boldsymbol{V}_d^{-1} \boldsymbol{X}_d$, so that $\boldsymbol{M} \approx 2(2 - r_d)\boldsymbol{B}$. Under this approximation, $\sigma_\Delta^2 \approx 2(2 - r)\boldsymbol{c}^T \boldsymbol{B}^{-1} \boldsymbol{c}$. For the $\boldsymbol{c}$ coefficients for the end-of-study pairwise contrast between $d$ and $d'$,

$$\boldsymbol{c}^T \boldsymbol{B}^{-1} \boldsymbol{c} \approx \frac{4 - 3\rho^2}{4} v_d - \frac{\rho^2}{2}\sqrt{v_d v_{d'}} + \frac{4 - 3\rho^2}{4} v_{d'}$$

Thus,

$$n = \frac{(2 - r)(z_q + z_{1-\alpha/2})}{2\Delta^2}\left((4 - 3\rho^2)(v_d + v_{d'}) - 2\rho^2 \sqrt{v_d v_{d'}}\right) \tag{15}$$

Notice also that if $\rho = 0$ then expression (15) would simplify to the cross-sectional equivalent, expression (14).